\documentclass{WileyMSP-template}
\usepackage{graphicx}
\usepackage{graphics}
\usepackage{dcolumn}
\usepackage{bm}
\usepackage{epsfig}
\usepackage{amsmath}
\usepackage{subfigure}
\usepackage{cases}
\usepackage{xcolor}
\usepackage{cite}
\usepackage{amssymb}

\begin{document}

\pagestyle{fancy}
\rhead{\includegraphics[width=2.5cm]{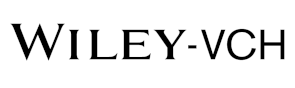}}

\title{Frequency-dependent squeezing via Einstein–Podolsky–Rosen entanglement based on silicon nitride microring resonators}

\maketitle


\author{Haodong Xu}
\author{Zijun Shu}  
\author{Nianqin Li}
\author{Yang Shen}
\author{Bo Ji}
\author{Yongjun Yang}
\author{Tengfei Wu}
\author{Mingliang Long}
\author{Guangqiang He*}%


\dedication{}

\begin{affiliations}
Haodong Xu, Zijun Shu, Nianqin Li, Yang Shen, Bo Ji, Prof. Guangqiang He\\
State Key Laboratory of Advanced Optical Communication Systems and Networks, Department of Electronic Engineering, 
Shanghai Jiao Tong University, Shanghai $\mathrm{200240}$, China\\
Email Address: gqhe@sjtu.edu.cn

Prof. Guangqiang He\\
State Key Laboratory of Precision Spectroscopy, East China Normal University, Shanghai $\mathrm{200062}$, China\\

Yongjun Yang, Tengfei Wu\\
Science and Technology on Metrology and Calibration Laboratory,
 Changcheng Institute of Metrology Measurement,
 Aviation Industry Corporation of China,
 Beijing $\mathrm{100095}$, China\\

Mingliang Long\\
Shanghai Astronomical Observatory, Chinese Academy of Sciences\\

\end{affiliations}



\begin{abstract}

Significant efforts have been made to enhance the performance of displacement sensors limited by quantum noise, such as gravitational wave detectors. Techniques like frequency-dependent squeezing have overcome the standard quantum limit in optomechanical force measurements, leading to substantial overall progress. These advancements, coupled with major developments in integrated photonics, have paved the way for the emergence of integrated Kerr quantum frequency combs (QFCs). A platform has been established for designing EPR entangled quantum frequency combs  using on-chip silicon nitride microring resonators, enabling thorough analysis and optimization of entanglement performance, as well as effective noise reduction adjustments. This platform, incorporating the quantum dynamics of Kerr nonlinear microresonators, supports at least 12 continuous-variable quantum modes in the form of 6 simultaneous two-mode squeezed pairs (EPR entangled pairs). Additionally, by selecting the detection angle of the idler mode, a single-mode squeezed state is generated in the signal mode. Given the frequency-dependent nature of the detection angle, frequency-dependent squeezing is achieved. A comparative analysis of the results under different dispersion conditions is also conducted.

\end{abstract}


\section{Introduction}
Quantum noise has become a critical limiting factor in precision displacement measurements, such as in gravitational wave detection \cite{yu2020nature,acernese2020virgo}. Quantum noise, driven by vacuum fluctuations entering the interferometer through the dark readout port \cite{caves1981quantum}, is divided into quantum back-action noise (photon radiation pressure noise) and shot noise. In gravitational wave detection, shot noise is typically driven by phase fluctuations of the incoming light field, while radiation pressure noise is driven by amplitude fluctuations. Using squeezed light reduces the interferometer's quantum noise by compressing the uncertainty in the electromagnetic field  \cite{walls1983squeezed,breitenbach1997measurement,schnabel2017squeezed}. Squeezing in a single field quadrature, independent of  frequency, has been used to reduce shot noise in observatories \cite{ligo2011gravitational, ligo2013enhanced}. By reflecting frequency-independent squeezed light out of a low-loss narrowband filter cavity, the optimal frequency-dependent squeezing angle can be achieved \cite{kimble2001conversion, chelkowski2005experimental, khalili2010optimal}. However, even with the highest quality cavity mirrors operating in ultra-high vacuum, an appropriate filter cavity must be 100 meters long to achieve the required linewidth and loss specifications for gravitational wave detection, which is evidently costly and challenging \cite{barsotti2018squeezed}.

To effectively reduce quantum back-action noise and overcome the standard quantum limit (SQL), various quantum non-demolition measurement techniques have been proposed, such as variational readout \cite{kimble2001conversion, vyatchanin1995frequency}, stroboscopic measurements \cite{braginsky1980science, vasilakis2015natphys}, two-tone measurements \cite{hertzberg2010natphys, suh2014science, shomroni2019natcomm}, and the optical spring effect \cite{chen2011genrelgrav}. Additionally,  methods have been proposed to generate frequency-dependent squeezed states without the need for external filter cavities, by leveraging the frequency-dependent characteristics of EPR entangled states to achieve optimal detuning \cite{ma2017proposal}. Other researchers have generated two-mode frequency-state squeezed vacuum via EPR entanglement \cite{yap2020generation, sudbeck2020demonstration} and controlled the squeezing angle using an auxiliary coherent locking field (CLF) \cite{vahlbruch2006coherent, chua2011backscatter}. Some have utilized detuned optical parametric oscillators to generate frequency-dependent squeezing \cite{junker2022frequency}, whose frequency-dependent Wigner function \cite{leonhardt1997measuring}
 was reconstructed through quantum tomography and exhibited rotation. These studies have shown promising noise reduction effects.  Our research, on the other hand, is based on silicon nitride ($\text{Si}_3\text{N}_4$) microring resonators, which generate EPR entangled pairs and further analyze frequency-dependent squeezing.

An optical parametric oscillator (OPO) is a nonlinear optical device that generates coherent light at different frequencies through the process of optical parametric amplification. In the aforementioned studies, it is commonly used to produce EPR entangled pairs that include signal and idler modes. In nondegenerate OPO, squeezing is a manifestation of the entanglement between the sidebands  \cite{hage2010towards}, and thus, bipartite entanglement criteria can be employed \cite{li2024platform} to deeply analyze frequency-dependent squeezing. Compared to the optical parametric oscillators  composed of nonlinear crystals (PPKTP) used in the aforementioned studies, our integrated microcavity features controllable dispersion and on-chip miniaturization. On-chip quantum frequency combs (QFCs) surpass the limitations of traditional free-space OPOs in terms of operability, integration, and scalability. When a Kerr resonator is weakly pumped, the resonator modes below the optical parametric oscillation  threshold are populated in pairs through a spontaneous parametric process. Waveguide has a small effective cross-sectional area \cite{dudley2006supercontinuum} and good scalability, allowing the formation of multiple two-mode squeezed pairs. Silicon nitride is compatible with CMOS technology and offers excellent properties such as low loss and a wide transparency window. The characteristics of the optical frequency comb generation process in microresonators are determined by the cavity waveguide dispersion. By utilizing advanced CMOS technology, we can precisely design the microresonator's ring-bus coupling rate and dispersion through cavity structure engineering.

In our work, we develop a platform based on integrated silicon nitride micro-ring resonators for dispersion and coupling engineering, successfully generating  EPR entangled frequency combs with at least 12 channels (six pairs), and each pair can be used to generate frequency-dependent single-mode squeezed state. Additionally, we can describe the full panorama of entanglement distribution across various modes using bipartite entanglement criteria.  We employ two different silicon nitride micro-ring cavity structures to correspond to normal and anomalous dispersion, respectively. We specifically analyze frequency-dependent squeezing via EPR entanglement, determining the readout angles that achieve maximum squeezing at different observation frequencies, and ultimately establish the relationship between entanglement bandwidth, threshold power, and the intrinsic quality factor $Q_0$. This approach allows for a precise control over quantum noise, may enhance the sensitivity of displacement measurement, such as gravitational wave detectors, and has the potential to expand the use of fully integrated entangled resources in quantum metrology.

The structure of this paper is as follows. Section \ref{sec2} provides an overview of the simulation model for the integrated silicon nitride microresonator and introduces dispersion and coupling engineering. Section \ref{sec3} proposes a theoretical model of four-wave mixing in the microresonator based on OPO theory, establishes the quantum entanglement criteria for signal and idler modes, and uses EPR entangled pairs to generate single-mode squeezed states. Section \ref{sec4} extracts and implements the necessary variables in the simulation process, examining the impact of these variables on the degree of entanglement and noise level. It provides a detailed analysis of entanglement characteristics and frequency-dependent squeezing, comparing the differences in various parameters and entanglement characteristics under normal and anomalous dispersion. Finally, the relationship curves between the intrinsic quality factor $Q_0$, entanglement bandwidth, and threshold power are obtained. Section \ref{sec5} summarizes the research findings, showcasing the process of analyzing the frequency dependence of squeezing based on a EPR entangled quantum frequency comb designed by silicon nitride microring resonators under different dispersion conditions.

\section{\label{sec2}Simulation model of microring resonators}
The pump light in the bus waveguide resonantly couples into the ring waveguide due to constructive interference, which allows spontaneous four-wave mixing (SFWM) to generate a quantum optical frequency comb (QFC). Figure 1 illustrates this third-order nonlinear process that satisfies energy conservation. The input continuous-wave pump light ($\Omega_p$) undergoes four-wave mixing, producing light signals ($\Omega_s$) and idlers ($\Omega_i$) with redistributed mode energies.
\begin{equation}\label{omega_relation}
    2\Omega_p = \Omega_s + \Omega_i.
\end{equation}
Typically, momentum conservation is expressed by the following equation:
\begin{equation}\label{omega_relation}
     2\vec{k}_p - \vec{k}_s - \vec{k}_i = \vec{0}.
\end{equation}

Figure 2(a) shows a typical structure of a microresonator with an additional coupling structure, where the OPO theory is applicable. Our on-chip microring resonator structure consists of a bus waveguide and a ring waveguide, with silicon nitride rib waveguides embedded in a silicon dioxide cladding. Figure 2(b) shows the coupling region and its input-output schematic. Since the optical field propagates along the x-axis in Figure 2(a), the effective cross-sectional area $A_{\text{eff}}$ of the ring waveguide \cite{dudley2006supercontinuum} is represented by Equation (3). It reflects the degree  of confinement of the optical field as it propagates within the resonator.
 The smaller the effective mode area, the stronger the confinement of the optical field by the resonator.

\begin{equation}\label{Aeff}
	A_{\mathrm{eff}}=\frac{\left(\iint_{-\infty}^{+\infty}|F(y, z)|^{2} d y d z\right)^{2}}{\iint_{-\infty}^{+\infty}|F(y, z)|^{4} d y d z}.
\end{equation}
Where $ \mathit{F(y,z)} $ represents the modal distribution in silicon nitride  and silicon dioxide, we assume that the modal distribution in the resonator is time-invariant.

The geometry of the coupling region is closely linked to the ring-bus coupling rate, allowing for easy extraction of the coupling rate, which reflects the input-output relationship of the resonator.

Here, we only consider the fundamental TE mode. The Lorentzian envelope of a cavity resonance can be observed (as shown in Figure 3, shaded area). Figure 3(b) and Figure 3(c) are enlarged subfigures under anomalous and normal dispersion conditions, respectively. Due to the frequency dependence of the material refractive index $n(\omega)$, the resonances vary across the spectrum.

To adjust the cavity variables such as detuning, dispersion, coupling, and loss rates through our design platform, theoretical preparations are required.

\subsection{Detuning and dispersion}

In this subsection, we discuss the nature of detuning and dispersion. We introduce the relative mode number $l$ ($l \in \mathbb{Z}$) to define the state modes adjacent to the pump mode $\omega_0$ ($l = 0$). A Taylor expansion analysis is applied to the resonant modes near $\omega_0$:
\begin{equation}\label{Taylor}
	\omega_l=\omega_0+D_1l+\frac{D_2}{2}l^2+\cdots=\omega_0+\sum_{\mathrm{n}=1}^{\infty} D_{\mathrm{n}}\frac{l^\mathrm{n}}{\mathrm{n}!}.
\end{equation}
\( D_1 = 2\pi \nu_f \), where \( \nu_f \) is the free spectral range (FSR). \( D_2 \) is related to group velocity dispersion (GVD) \cite{shi2023entanglement}, where \( D_2 > 0 \) (or \( D_2 < 0 \)) indicates anomalous (or normal) dispersion. Here, we neglect higher-order dispersion, so we set \( D_n = 0 \) when \( n \geq 3 \).
 We can define the integrated dispersion \( D_{\text{int}} = \omega_l - \omega_0 - D_1 l \) \cite{godey2014stability}, which can be well approximated by a quadratic polynomial around \( \omega_0 \). We perform a comparative analysis of the anomalous and normal dispersion in silicon nitride microring resonators.

We assume that the cavity maintains the same temperature throughout the entire microring structure at any given moment. The resonant modes are given by $\omega_l$. 
Since the comb lines of QFCs are always equidistant,  we can obtain the following equation:
\begin{equation}\label{A2}
	\Omega_l=\Omega_0+D_1l.
\end{equation}
When we ignore higher-order small terms,  the resonance modes  is:
\begin{equation}\label{A3}
	\omega_l=\omega_0+D_1l+\frac{D_2}{2}l^2.
\end{equation}
Consider Eq.\eqref{A2}\eqref{A3}, the normalized cold cavity detuning $\Delta_c$ at mode $l$ can be expressed by the following eqution:
\begin{equation}\label{A4}
	\begin{aligned}
	\Delta_c{}_,{}_l &=\omega_l-\Omega_l\\
	&=\sigma_c+\frac{D_2}{2} l^2,
	\end{aligned}
\end{equation}
where  $\sigma_c =\omega_0-\Omega_0$, the normalized cold cavity pump detuning. Thus, the relationship between detuning and dispersion  is established.

\subsection{\label{GAL}Coupling, loss and gap }

In this section, we analyze the relationship between the input-output parameters, namely the coupling ($\gamma$) and loss rate ($\mu$) of the cavity, and the  cavity parameters: $\kappa$ and quality factor. We assume that the resonator loss is represented as an effective phantom channel \cite{vernon2015spontaneous,vernon2015strongly}, and the beam splitter transmission characteristics are used to ensure $\gamma \ll \nu_f$ and $\mu \ll \nu_f$. As illustrated in Fig. 2(b),  $\gamma = |t_2|^2\nu_f = (1 - |t_1|^2)\nu_f$ and $\mu$ is simplified to $\mu = \alpha L \nu_f$, where $L = 2 \pi R$. $R = \frac{D_\text{in} + D_\text{out}}{4}$, the radius of the microring resonator, where $D_\text{in}$ and $D_\text{out}$ are the inner and outer diameters of the microring resonator, respectively. Additionally, $\alpha \approx f_0 / (Q_0 \cdot R \cdot \nu_f)$ is the absorption coefficient (m$^{-1}$), where $f_0$ (Hz) is the resonance frequency.

Next, we set $b_2 = 0$ and adjust the distance between the ring waveguide and the bus waveguide (gap) to scan the transmission from the added port to the through port ($\left| t_1 \right|^2$). The ratio $r = \gamma / \mu$ indicates the relationship between coupling and internal loss. When $r < 1$, it signifies under-coupling; $r > 1$ indicates over-coupling; and $r = 1$ represents critical coupling. In our simulation, we use  over-coupled structures that, while leading to lower intracavity power, allow for more efficient power extraction from the resonator.

For resonators used to generate quantum optical frequency combs, the quality factor 
Q is a crucial parameter. Depending on the Q factor, the resonator can achieve exceptionally high field enhancement.  The Q value reflects the microcavity's ability to store optical energy and can be expressed as:
\begin{equation}
Q = \omega_0 \tau_p = \frac{\omega_0}{\kappa},
\end{equation}
where \( \omega_0 \) is the central frequency of the resonance peak, $\tau_p$ is the photon lifetime and $\kappa$  is the full width at half maximum (FWHM) of the resonance peak \cite{ji2024simultaneous}.
Meanwhile, the value of the loss coupling rate (rad/s) $\kappa_0=\omega_0/Q_0\approx c\alpha/n_g= \mu$, where $n_g$ is the group refractive index of silicon nitride around $\Omega_0$.  The total loss rate $\kappa=\kappa_0(1+r)= \Gamma$ and the ring-bus coupling rate $\kappa_{ex}=\kappa-\kappa_0=\omega_0/Q_{ex}= \gamma$, where $Q_{ex}$ is the external quality factor \cite{javid2023chipscale}. And the total quality factor $Q$ satisfies:
\begin{equation}
		\frac{1}{Q} = \frac{1}{Q_0}+\frac{1}{Q_{ex}}.
\end{equation}
Thus, we can relate the gap to $Q_{ex}$ using the ratio $r$.

\subsection{Actual simulation}

 Our goal is to design a dispersion-flattened waveguide  to minimize phase mismatch in the FWM process. By adjusting the parameters of the silicon nitride integrated microcavity structure, such as geometry, size, and selecting nonlinear materials, the dispersion characteristics  are tuned to match the phase mismatch. To generate more operationally entangled state pairs, dispersion engineering is crucial. By fitting \(D_{\text{int}}\) to near zero over the widest spectral range, optimal  phase matching conditions can be ensured.
\begin{equation}\label{pm}
	\Delta k = \frac{2\omega_{p}n(\omega_{p})}{c}-\frac{\omega_{s}n(\omega_{s})}{c}-\frac{\omega_{i}n(\omega_{i})}{c}.
\end{equation}

We selected two microcavity structures for the cases of anomalous dispersion and normal dispersion, which can generate classical optical solitons. Fig.4(a) shows the micro-ring structure under anomalous dispersion, where the radius is $R=23\,\mu\text{m}$, 
the waveguide width and bus width are $W_{R1}=W_{B1}=1610\,\text{nm}$, the height is $H_1=h_1=800\,\text{nm}$, and the angle $\theta=90^\circ$. For this structure, the simulation results show that the free spectral range $\nu_f$ is $989.592\,\text{GHz}$, the frequency of mode 0 $f_0$ is $193.251\,\text{THz}$, the second-order dispersion coefficient $D_2$ is $1.435\times 2\pi\times 10^7\,\text{rad/s}$, and the effective area $A_\text{eff}$ is $1.10\,\mu\text{m}^2$. Fig.4(b) shows the micro-ring structure under normal dispersion, where the radius is $R=23\,\mu\text{m}$, the waveguide width and bus width are $W_{R2}=W_{B2}=1710\,\text{nm}$, the height is $H_2=h_2=400\,\text{nm}$, and the angle $\theta=90^\circ$. In this configuration, $\nu_f$ is $1019.553\,\text{GHz}$, $f_0$ is $193.797\,\text{THz}$, $D_2$ is $-5.676\times 2\pi\times 10^8\,\text{rad/s}$, and $A_\text{eff}$ is $0.968\,\mu\text{m}^2$.

Using a fifth-order polynomial fit, we simulated the dispersion of the silicon nitride ring with a radius of $23\,\mu\text{m}$ and oxide coating, obtaining parameters that fit well with $D_\text{int}$, as shown by the dispersion curves in Fig.4(c) and Fig.4(d). We assume the intrinsic quality factor $Q_0=10^6$, leading to $\kappa_0=1.21\times10^9$. By setting the coupling ratio $r=\gamma/\mu=1.222$ (overcoupling), we determine that the gap is $490\,\text{nm}$.

\section{\label{sec3}EPR entanglement dynamics}
In this section, we will expound the quantum dynamics in the resonator based on Sec. \ref{sec2}.
\subsection{Hamiltonian}
The system Hamiltonian can characterize the optical nonlinear processes in the quantum mechanics domain. Each resonant mode is described by an annihilation operator \(\hat{a}_j\), where \(j = p, s, i\). In a system undergoing a four-wave mixing process, the total energy of the system can be divided into two parts: the free Hamiltonian \(\hat{H}_0\) and the nonlinear Hamiltonian \(\hat{H}_{NL}\), which are expressed as follows:
\begin{equation}
\hat{H} = \hat{H}_0 + \hat{H}_{NL}.
\end{equation}
The free Hamiltonian $\hat{H_{0}}$ is given by:
\begin{eqnarray}
	\hat{H_{0}}=\hbar\sum_{j = p, s, i} \omega_{j} \hat{a}_{j}^{\dagger} \hat{a}_{j}.
\end{eqnarray}
Generally, the nonlinear Hamiltonian for the four-wave mixing process \cite{zeng2024quantum,wang2004broadband} can be written as:
\begin{equation}
\hat{H}_{\text{NL}} = \hat{H}_{\text{SPM}} + \hat{H}_{\text{XPM}} + \hat{H}_{\text{FWM}},
\end{equation}
where \(\hat{H}_{\text{SPM}}\), \(\hat{H}_{\text{XPM}}\), and \(\hat{H}_{\text{FWM}}\) represent the self-phase modulation (SPM), cross-phase modulation (XPM), and four-wave mixing (FWM) processes, respectively. The SPM, XPM and FWM  terms affect the oscillation process as well as the system's noise and entanglement properties. The four-wave mixing process is crucial for energy transfer among the four modes. The formula for the nonlinear Hamiltonian \(\hat{H}_{\text{NL}}\) can be specifically written as:
\begin{equation}
\begin{aligned}
\hat{H}_{\text{NL}} = & -\hbar \eta \left[ \frac{1}{2} \left( \hat{a}_p^\dagger \hat{a}_p^\dagger \hat{a}_p \hat{a}_p + \hat{a}_s^\dagger \hat{a}_s^\dagger \hat{a}_s \hat{a}_s + \hat{a}_i^\dagger \hat{a}_i^\dagger \hat{a}_i \hat{a}_i \right) \right. \\
& + 2 \left( \hat{a}_p^\dagger \hat{a}_s^\dagger \hat{a}_p \hat{a}_s + \hat{a}_p^\dagger \hat{a}_i^\dagger \hat{a}_p \hat{a}_i + \hat{a}_s^\dagger \hat{a}_i^\dagger \hat{a}_s \hat{a}_i \right) \\
& \left. + \left( \hat{a}_s^\dagger \hat{a}_i^\dagger \hat{a}_p \hat{a}_p + \hat{a}_p^\dagger \hat{a}_p^\dagger \hat{a}_s \hat{a}_i \right) \right].
\end{aligned}
\end{equation}

Here, the lower bound estimate of the nonlinear coupling coefficient is given by
$\eta=\hbar\omega_0^2cn_2/(n_0^2V_{\mathrm{eff}})$ \cite{chembo2016quantum},
representing the per photon frequency shift of the resonance due to the \(\chi^{(3)}\) nonlinearity. Here, \(c\) is the vacuum speed of light, and \(n_2\) is the nonlinear index of Si\(_3\)N\(_4\), which is associated with the refractive index \(n_0\). In our Si\(_3\)N\(_4\) microresonator, \(n_2 = 2.6 \times 10^{-19} \, \text{m}^2/\text{W}\). The effective mode volume \(V_{\text{eff}}\) can be defined as
\begin{equation}
	V_\text{eff}=\frac{\int n_0^2|F(x,y,z)|^2dV\int|F(x,y,z)|^2dV}{\int n_0^2|F(x,y,z)|^4dV}.
\end{equation}

When the WGM resonator is a microring resonator, the upper bound estimate of $V_\text{eff}$ can be approximated by $V_\text{eff} \approx A_\text{eff} \cdot 2 \pi R$.

\subsection{Heisenberg-Langevin equations}
The Heisenberg-Langevin equations \cite{whalen2016time} combine the Heisenberg equations with Langevin noise terms to describe the dynamics of an open system in the presence of quantum noise. The combined Heisenberg-Langevin equation is:
\begin{equation}
\begin{aligned}
\frac{d \hat{a}_j}{dt}& = -\frac{i}{\hbar} [\hat{a}_j, \hat{H}] - \kappa \sum_j \hat{a}_j + \sqrt{2\kappa_{ex}} \sum_j \hat{a}_{j}^{\mathrm{in}}\\
& +\sqrt{2\kappa_0} \sum_j \hat{a}_{j}^{\mathrm{loss}}, \quad j = p, s, i.
\end{aligned}
\end{equation}
We assume that the modes under consideration have similar field profiles and share the same total loss rate $\kappa$, intrinsic loss rate $\kappa_0$, and ring-bus coupling rate $\kappa_{ex}$, where $\kappa = \kappa_0 + \kappa_{ex}$. The annihilation operators $\hat{a}^{\mathrm{in}}$ and $\hat{a}^{\mathrm{loss}}$ represent the incident and loss modes of the resonator, respectively. The loss modes are assumed to be vacuum states, and the incident signal modes and idle modes are also vacuum states. The expectation of incident pump mode is represented by the following expression: 
\begin{equation}
\left\langle\hat{a}_{p}^{\mathrm{in}}(t)\right\rangle = A^{\text{in}} = \sqrt{\frac{P_{\text{in}}}{\hbar \Omega_0}},
\end{equation}
where $P_{\mathrm{in}}$ (Watt) is the pump laser power in the bus waveguide.

By applying the rotating wave approximation (RWA), where \(\hat{a}_j e^{-i \omega_j t}\) is used to replace \(\hat{a}_j\), the Heisenberg-Langevin equations for the pump, signal, and idle modes become

\begin{equation}\label{Heisenberg-Langevin equation}
    \left\{
    \begin{aligned}
        \frac{\mathrm{d} \hat{a}_{p}}{\mathrm{d} t} &= \mathrm{i} \eta \Big(\hat{a}_{p}^{\dagger} \hat{a}_{p} \hat{a}_{p} + 2 \hat{a}_{s}^{\dagger} \hat{a}_{s} \hat{a}_{p} + 2 \hat{a}_{i}^{\dagger} \hat{a}_{i} \hat{a}_{p} + 2 \hat{a}_{p}^{\dagger} \hat{a}_{s} \hat{a}_{i} \Big) \\
        &\quad - \kappa \hat{a}_{p} - \mathrm{i}  \sigma_c \hat{a}_{p} + \sqrt{2 \kappa_{ex}} \hat{a}_{p}^{\mathrm{in}} + \sqrt{2 \kappa_0} \hat{a}_{p}^{\mathrm{loss}}, \\
        \frac{\mathrm{d} \hat{a}_{s}}{\mathrm{d} t} &= \mathrm{i} \eta \Big( 2 \hat{a}_{p}^{\dagger} \hat{a}_{p} \hat{a}_{s} + \hat{a}_{s}^{\dagger} \hat{a}_{s} \hat{a}_{s} + 2 \hat{a}_{i}^{\dagger} \hat{a}_{i} \hat{a}_{s} + \hat{a}_{p}^{2} \hat{a}_{i}^{\dagger} \Big) \\
        &\quad - \kappa \hat{a}_{s} - \mathrm{i}  \Delta_{-l} \hat{a}_{s} + \sqrt{2 \kappa_{ex}} \hat{a}_{s}^{\mathrm{in}} + \sqrt{2 \kappa_0} \hat{a}_{s}^{\mathrm{loss}}, \\
        \frac{\mathrm{d} \hat{a}_{i}}{\mathrm{d} t} &= \mathrm{i} \eta \Big( 2 \hat{a}_{p}^{\dagger} \hat{a}_{p} \hat{a}_{i} + 2 \hat{a}_{s}^{\dagger} \hat{a}_{s} \hat{a}_{i} + \hat{a}_{i}^{\dagger} \hat{a}_{i} \hat{a}_{i} + \hat{a}_{p}^{2} \hat{a}_{s}^{\dagger} \Big) \\
        &\quad - \kappa \hat{a}_{i} - \mathrm{i}  \Delta_{+l} \hat{a}_{i} + \sqrt{2 \kappa_{ex}} \hat{a}_{i}^{\mathrm{in}} + \sqrt{2 \kappa_0} \hat{a}_{i}^{\mathrm{loss}},
    \end{aligned}
    \right.
\end{equation}
where  \(\Delta_{-l}\) and \(\Delta_{+l}\) are the normalized cold cavity detuning for the signal mode and idler mode, respectively. Here,  $\kappa_{ex}$ and \(\kappa_0\) are constant parameters that do not vary with the mode number \(l\).

\subsection{Steady-state equations}
We can apply linearization methods by expanding each field operator \(\hat{a}_j\) into its steady-state average value \(\alpha_j\) and a fluctuation operator \(\delta \hat{a}_j\), such that $\hat{a}_j = \alpha_j + \delta\hat{a}_j$ In the steady state, \(\alpha_j\) remains constant. Therefore, by setting \(\delta \hat{a}_j = 0\) and \(\frac{\mathrm{d} \alpha_j}{\mathrm{d} t} = 0\), we can derive the steady-state Heisenberg-Langevin equations. In this situation, the input fields for the signal and idle modes, as well as the losses for the signal, idle, and pump modes, are in the vacuum state, so:$\alpha_{s}^{\mathrm{in}}=\alpha_{i}^{\mathrm{in}}=\alpha_{j}^{\mathrm{loss}}=0$.

For simplicity, we take the phase of the external pump
as a reference.
 Thus we set $\alpha_{j}=A_{j} \mathrm{e}^{\mathrm{i}\theta_{j}}$, $\alpha_{p}^{\mathrm{in}}=A^{\mathrm{in}} \mathrm{e}^{\mathrm{i}\theta_{\mathrm{in}}}$, $\Theta=\theta_{s}+\theta_{i}-2\theta_{p}$, $\psi=\theta_{\mathrm{in}}-\theta_{p}$. For simplicity, we set $A_{s}=A_{i}=A$ and $\Delta_{+l}=\Delta_{-l}=\Delta$, with the scale to the cavity resonance half-width $\kappa/2$ omitted.The
external pump power is defined as 
$F=\sqrt{\frac{2\gamma \eta}{\hbar\Omega_0\Gamma^3}P_{\mathrm{in}}}$ , with $A^{\mathrm{in}}=F\sqrt{\frac{\Gamma^3}{2\gamma \eta}}$. Based on these variables, we get:
\begin{align}\label{steady-state solution}
    \left\{
    \begin{array}{ll}
        \quad A_p^4 &= 1 + (\sigma_c - \frac{D_3}{2} - 2A_p^2 - 3A^2)^2,  \\
        \quad F^2 &= A_p^2  (1 + 2  \frac{A^2}{A_p^2} )^2  \\
        &\quad +A_p^2[\sigma_c-A_p^2- 2 \frac{A^2}{A_p^2}(\sigma_c-\frac{D_3}{2}-3A^2)]^2 , \\
        \sin(\Theta) &= \frac{1}{A_p^2} ,\\
        \cos(\Theta) &=  \frac{1}{A_p^2} \left( \Delta  - 3A^2 - 2A_p^2 \right) ,\\
        \sin(\psi) &= \frac{A_p}{F} \left[ \sigma_c - A_p^2 -  2 \frac{A^2}{A_p^2} (-3A^2 - \frac{D_3}{2} + \sigma_c )\right] ,\\
        \cos(\psi) &= \frac{A_p}{F} \left( 1 + 2 \frac{A^2}{A_p^2} \right).
    \end{array}
    \right.
\end{align}
Eq. (\ref{steady-state solution}) contains  variables: ${A}_{p}$, ${A}$, ${A}^{in}$, ${\sigma_c}$, and ${\Delta}$. By assigning actual values to any two of these parameters, we can determine the numerical relationship between the remaining three variables. This implies that any two of the remaining three variables can be expressed as functions of the third one.

\subsection{Quantum fluctuation equations}
To explore the quantum properties of the signal and
idler, we need the quantum fluctuation equations derived
from Eq.(\ref{Heisenberg-Langevin equation}). Since we have already calculated the steady-state equations, we can obtain the quantum fluctuation equations by eliminating the steady-state terms from Eq.(\ref{Heisenberg-Langevin equation}). We treat the pump field as a classical beam, thus $\delta\hat{a}_{p}=0$. Also, high order fluctuations are neglected.

Define a vector containing the fluctuations of the signal and idler as:
\begin{equation}
	\delta \hat{\mathbf{A}}=\left(\delta \hat{a}_{s} \mathrm{e}^{-\mathrm{i} \theta_{s}}, \delta \hat{a}_{s}^{\dagger} \mathrm{e}^{\mathrm{i} \theta_{s}}, \delta \hat{a}_{i} \mathrm{e}^{-\mathrm{i} \theta_{i}}, \delta \hat{a}_{i}^{\dagger} \mathrm{e}^{\mathrm{i} \theta_{i}}\right)^{\mathrm{T}},
\end{equation}
where $\theta_{j}$ is the phase of the mean value $\alpha_{j}=A_{j}\mathrm{e}^{\mathrm{i}\theta_{j}}$. The time evolution of these fluctuations are given by:
\begin{equation}
	\frac{\mathrm{d} \delta \hat{\mathbf{A}}}{\mathrm{d} t}=M_{a} \cdot \delta \hat{\mathbf{A}}+T_{a}^{\mathrm{in}} \cdot \delta \hat{\mathbf{A}}^{\mathrm{in}}+T_{a}^{\mathrm{loss}} \cdot \delta \hat{\mathbf{A}}^{\mathrm{loss}},
\end{equation}
where $T_{a}^{\mathrm{in}}=\mathrm{diag}\left(\sqrt{2{\kappa_{ex}}}, \sqrt{2{\kappa_{ex}}}, \sqrt{2{\kappa_{ex}}}, \sqrt{2{\kappa_{ex}}}\right)$, $T_{a}^{\mathrm{loss}}=\mathrm{diag}\left(\sqrt{2{\kappa_0}}, \sqrt{2{\kappa_0}}, \sqrt{2{\kappa_0}}, \sqrt{2{\kappa_0}}\right)$. The matrix $M_a$  is derived from the linearization process, with its elements related to the average values of the fields and the detuning.

The evolution of these fluctuations in the frequency domain can be described using the Fourier transform. After incorporating the input-output relationship of the resonator 
\begin{equation}
	\hat{a}^{\mathrm{out}}=-\hat{a}^{\mathrm{in}}+\sqrt{2\kappa_{ex}} \hat{a},
\end{equation} we obtain:
\begin{equation}
	\begin{aligned}
		\delta \hat{\mathbf{A}}^{\mathrm{out}}(\omega) &=-\delta \hat{\mathbf{A}}^{\mathrm{in}}+T_{a} \delta \hat{\mathbf{A}}\\
		&=[T_{a}\left(\mathrm{i} \omega I-M_{a}\right)^{-1} T_{a}^{\mathrm{in}}-I] \cdot \delta \hat{\mathbf{A}}^{\mathrm{in}}\\
  &+T_{a}\left(\mathrm{i} \omega I-M_{a}\right)^{-1} T_{a}^{\mathrm{loss}}\cdot \delta \hat{\mathbf{A}}^{\mathrm{loss}},
	\end{aligned}
\end{equation}
where $T_{a}=\mathrm{diag}\left(\sqrt{2\kappa_{ex}}, \sqrt{2\kappa_{ex}}, \sqrt{2\kappa_{ex}}, \sqrt{2\kappa_{ex}} \right)$,
and $I$ is the identity matrix.

Therefore, the output spectral noise density matrix is defined by:
\begin{equation}
	\begin{aligned}
		S_{a}(\omega)&=\left\langle\delta \hat{\mathbf{A}}^{\mathrm{out}}(\omega) \delta \hat{\mathbf{A}}^{\mathrm{out} ,\mathrm{T}}(-\omega)\right\rangle \\
		&=[T_{a}\left(\mathrm{i} \omega I-M_{a}\right)^{-1} T_{a}^{\mathrm{in}}-I]\cdot M_c \\ 
  &\cdot [T_{a}\left(\mathrm{-i} \omega I-M_{a}\right)^{-1} T_{a}^{\mathrm{in}}-I]^T
		+T_{a}\left(\mathrm{i} \omega I-M_{a}\right)^{-1}\\
  &\cdot T_{a}^{\mathrm{loss}}\cdot M_c\cdot [T_{a}\left(\mathrm{-i} \omega I-M_{a}\right)^{-1}T_{a}^{\mathrm{loss}}]^T,
	\end{aligned}
\end{equation}
where the matrix $M_c=\left(\begin{array}{cccc} 0 & 1 & 0 & 0 \\0 & 0 & 0 & 0 \\0 & 0 & 0 & 1 \\0 & 0 & 0 & 0\end{array}\right)$.

\subsection{EPR  and bipartite entanglement criterion }
To quantitatively analyze the entanglement between signal and idler, we apply the criterion from Ref.\cite{duan2000inseparability} to determine the degree of entanglement  $C_s$.  
We define the amplitude quadrature \(\hat{x}_j\) (\(j = s, i\)) and phase quadrature \(\hat{y}_j\) (\(j = s, i\)) operators \cite{wang2023ultrasensitive}, which are functions of \(\hat{a}_j\) and \(\hat{a}_j^\dagger\):
\begin{equation}
\hat{x}_j = \frac{\hat{a}_j + \hat{a}_j^\dagger}{\sqrt{2}}, \quad \hat{y}_j = \frac{-i\hat{a}_j + i\hat{a}_j^\dagger}{\sqrt{2}}.
\end{equation}
By rotating the detection angles of the signal and idler beam ($\theta_s$ , $\theta_i$), we can obtain:
\begin{equation}
	\left(\delta \hat{x}_{s}, \delta \hat{x}_{i}, \delta \hat{y}_{s}, \delta \hat{y}_{i}\right)^{\mathrm{T}}=P\left(\delta \hat{a}_{s}, \delta \hat{a}_{s}^{\dagger}, \delta \hat{a}_{i}, \delta \hat{a}_{i}^{\dagger}\right)^{\mathrm{T}},
\end{equation}
where P =$\dfrac{1}{\sqrt{2}}\begin{pmatrix}
e^{-i\theta_s} & e^{i\theta_s} & 0 & 0 \\
0 & 0 & e^{-i\theta_i} & e^{i\theta_i} \\
-ie^{-i\theta_s} & ie^{i\theta_s} & 0 & 0 \\
0 & 0 & -ie^{-i\theta_i} & ie^{i\theta_i}
\end{pmatrix}$.

Introduce the sum and subtraction basis:

\begin{equation}
	\hat{x}_{\pm}=\frac{\hat{x}_{s} \pm \hat{x}_{i}}{\sqrt{2}}, 
	\quad\hat{y}_{\pm}=\frac{\hat{y}_{s} \pm \hat{y}_{i}}{\sqrt{2}},
\end{equation}
and we obtain the fluctuation vector
\begin{equation}
	\delta \hat{X}_{\pm}=\left(\delta \hat{y}_{+}, \delta \hat{x}_{+}, \delta \hat{y}_{-}, \delta \hat{x}_{-}\right)^{\mathrm{T}}
 =Q\left(\delta \hat{x}_{s}, \delta \hat{x}_{i}, \delta \hat{y}_{s}, \delta \hat{y}_{i}\right)^{\mathrm{T}},
\end{equation}
where Q =$\dfrac{1}{\sqrt{2}}\begin{pmatrix}
0 & 0 & 1 & 1 \\
1 & 1 & 0 & 0 \\
0 & 0 & 1 & -1 \\
1 & -1 & 0 & 0
\end{pmatrix}$.

The spectral noise density matrix $S_{\hat{X}_{\pm}}(\omega)$ is calculated by:
\begin{equation}
		S_{\hat{X}_{\pm}}(\omega)=\left\langle\delta \hat{X}_{\pm}(\omega) \delta \hat{X}_{\pm}^{\mathrm{T}}(-\omega)\right\rangle
		=Q\cdot P\cdot S_{a}(\omega) \cdot (Q\cdot P)^{\mathrm{T}},
\end{equation}
where $S_a(\omega)$ is defined in Eq.(24).

The Duan criterion has the following form \cite{gonzalez2017third}:
\begin{equation}\label{criterion}
	C_s =(\Delta \hat{x}_{-})^{2}+(\Delta \hat{y}_{+})^{2}-|G|\geq 0,
\end{equation}
where $(\Delta \hat{x}_{-})^{2}=S_{\hat{X}_{\pm}}(\omega)(4,4)$,  $(\Delta \hat{y}_{+})^{2}=S_{\hat{X}_{\pm}}(\omega)(1,1)$, and $G=\cos(\theta_{s}-\theta_{i})$. If the Duan criterion is not satisfied, i.e., \( C_s < 0 \), then the bipartite elements are entangled. The smaller the value of \( C_s \), the better the quantum entanglement.

The condition for maximizing the degree of entanglement obtained through simulation corresponds to the condition for the best quadrature of the single-mode squeezed state of the signal mode, which is also the condition for minimizing quantum back-action noise to the greatest extent. Thus, we can utilize this EPR entangled QFCs platform to achieve excellent noise reduction.

As shown in Fig. 5, the signal and idler beams have EPR entangled sidebands. Initially, the quantum statistics of both beams follow a thermal state distribution, resulting in considerable quantum noise. When we detect the idler beam at a specific angle \(\theta_i\), the quantum statistics of the signal beam are instantaneously squeezed at another angle \(-\theta_i\) \cite{ma2017proposal}, allowing us to overcome the Standard Quantum Limit (SQL). We define the readout angle \(\phi = \theta_s - \theta_i\), which can be used to further study the frequency-dependent squeezing via EPR entanglement in relation to the readout angle and observation frequency.


\section{\label{sec4}Entanglement and squeezing analysis}
In this section, we conduct a comprehensive study of signal-idler two-mode entanglement based on simulation results derived from practical parameters. By utilizing the two-mode squeezed state, we generate a single-mode squeezed state in the signal mode, which is then analyzed for its frequency-dependent squeezing characteristics. Additionally, a comparative analysis is carried out to evaluate the entanglement properties under both normal and anomalous dispersion conditions.
\subsection{\label{sec4A}Simulation process}

Our entire simulation process is as follows. First, the structural parameters of the microring resonator need to be provided to simulate the effective refractive index $n_{\mathrm{eff}}(\omega)$. Second, we determine the center frequency of the pump light ($\Omega_0\approx$ 193 THz$\times2\pi$) and calculate the resonance mode and $D_{\mathrm{int}}$. Next, we simulate the coupling relationship between the microring resonator and the bus waveguide, which allows us to extract parameters such as $\kappa_{ex}$. Then, we evaluate the entanglement relationship between the signal and idler by using the criterion of $C_s<0$ as a signature of bipartite entanglement,and analyze frequency-dependent squeezing. When the comb generation process approaches the OPO threshold, the linearization method mentioned in Section 3 becomes invalid, and higher-order fluctuations will significantly affect the entanglement properties \cite{ng2023quantum,vendromin2024quantum}. Therefore, in the entanglement analysis of this section, we only consider scenarios far from the threshold. Finally, we analyze the relationship between the intrinsic quality factor \(Q_0\) and the entanglement bandwidth and threshold power.

\subsection{EPR entangled QFCs}
Based on our design platform, we successfully generate EPR entangled frequency combs with 12 channels (6 pairs), as depicted in Fig. 6. Assume that  \(\sigma_c = 8 \, \text{GHz}\) for anomalous dispersion and \(\sigma_c = 18 \, \text{GHz}\) for normal dispersion.
Our QFC is under the influence of hysteresis \cite{chembo2010modal}.

In Fig. 6, the S-shaped brown curve represents all frequency modes below the OPO threshold (spontaneous FWM process), while the colorful bifurcation structure indicates all frequency modes above the OPO threshold (stimulated FWM process). Fig.6 (a) depicts the case of anomalous dispersion, and Fig.6 (b) illustrates the case of normal dispersion.
Specifically, consider the case where $l=4$. The solutions to Eq. (20) are illustrated in Fig. 7. In stages I and IV, the orange and red curves indicate that the comb generation system is below the OPO threshold. Here, the amplitudes of the signal and idler modes are zero, while only the intracavity pump mode amplitude changes with the injected pump amplitude. In stages II and III, the blue and green curves show that the comb generation system is above the OPO threshold, and all intracavity mode amplitudes vary in response to the injected pump power. Although the brown curve also represents the system below the OPO threshold, the solutions in these regions correspond to unstable values that are not relevant for our simulation.

The result of phase modulation beyond the OPO threshold is evident, as the amplitude of the cavity modes fluctuates in response to the injected pump power. This phenomenon can be attributed to the phase modulation that occurs when the pump power increases to trigger the OPO.  Furthermore, the specific characteristics of the relationship between the intracavity modes and the injected pump mode are detailed in Ref. \cite{matsko2005optical}.

Then, we take the fourth and third mode ($l$=4, $l$=3) into consideration, which can be analyzed in four distinct stages, shown in Fig. 8(a) and Fig. 9(a). Stages \uppercase\expandafter{\romannumeral1} and \uppercase\expandafter{\romannumeral4} denote a QFC that is below the threshold, and it is noteworthy that these two stages exhibit bistability, which means each stage converges to either stage depending on the detuning procedure employed. Correspondingly, Stages \uppercase\expandafter{\romannumeral2} and \uppercase\expandafter{\romannumeral3} indicate a QFC that is above the threshold. By rotating different angles $\theta_{s}$ and $\theta_{i}$ to locate the minimum value of $C_s$, we can obtain a panoramic view of the QFC entanglement distribution. Fig. 8(b) and Fig. 9(b) showcases the entanglement distribution in the injected pump amplitude $A^{in}$ and observation frequency f regime, assuming the coupling rate $r=1.222$ and pump detuning $\sigma_c = 8$ GHz. Similarly,  Fig. 8(c) and Fig. 9(c) show the four stages under normal dispersion, while  Fig. 8(d) and Fig. 9(d) illustrate the entanglement distribution, assuming the coupling rate $r=1.222$  and pump detuning $\sigma_c = 18$ GHz. We observe that as the injected pump amplitude  increases, the peak value of the entanglement degree also rises in stage I. Our simulation results show that there is only one minimum in stages I and IV, while stages II and III have two minima.
By comparing the entanglement distribution diagrams under anomalous and normal dispersion, we can observe that in the case of normal dispersion, the optimal observation frequency points are overall shifted upwards. By adjusting the injected pump amplitude and observation frequency, we can achieve entanglement extremum and maximize noise reduction.
\subsection{Frequency-Dependent Squeezing}
To replace narrowband filter cavities and achieve cost-effective and efficient frequency-dependent squeezing, we can utilize each of the six two-mode squeezed states generated in the aforementioned simulation to prepare a single-mode squeezed state of the signal mode (this paper presents simulation results for \( l = 4 \)). For frequency-dependent squeezing, low frequencies primarily exhibit amplitude squeezing, while high frequencies show phase squeezing. When we detect the idler mode at an angle \(\theta_i\), the signal mode is instantaneously squeezed at an angle of \(-\theta_i\). By analyzing the relationship between the entanglement degree \(C_s\), the observation frequency f, and the readout angle \(\phi\) (as shown in Fig. 10), we can quantify the frequency-dependent squeezing. For a given observation frequency, we can determine the corresponding optimal readout angle. The silicon nitride microring resonator we use features an integrated design, offering advantages of compact size and high performance in applications. This research can improve the sensitivity of displacement sensors such as gravitational wave detectors and be applied in the field of quantum precision measurement.

\subsection{Entanglement bandwidth and threshold power}
In this subsection, we will highlight the differences between anomalous dispersion and normal dispersion.

First, we explore the relationship between the intrinsic quality factor $Q_0$, entanglement bandwidth $\delta f$, and threshold power $P_\text{th}$. Fig. 11 illustrates the curves for both anomalous and normal dispersion conditions. It is evident that, within an appropriate range, $Q_0$ is inversely proportional to the entanglement bandwidth $\delta f$ and positively correlated with the threshold power $P_\text{th}$. Additionally, we observe a slight increase in the entanglement extremum as $Q_0$ increases. Here, we fix the coupling rate $r=1.222$. When calculating the entanglement bandwidth, we use a mode number of 4, specifying the pump amplitude $A^\text{in}$ and detuning, and select the fourth stage among four stages that best represents the entanglement bandwidth, using $1/e$ of the entanglement extremum as the boundary (where $e$ is the natural constant).

We can see that, for the same $Q_0$ and detuning, the threshold power for normal dispersion is slightly lower than that for anomalous dispersion. This is because, in our simulations, the nonlinear coefficient $\eta_1$ for normal dispersion is greater than $\eta_2$ for anomalous dispersion ($\eta_1=27.75$, $\eta_2=20.93$). This means that, for the same optical power, normal dispersion exhibits stronger nonlinear effects and therefore requires a lower threshold power.

Unlike the threshold power required for the earliest comb tooth to appear \cite{chembo2010modal}, the threshold power we refer to here is the minimum pump power required to grow comb teeth with a specific mode number (with $l=1$ selected in this paper). As $Q_0$ increases, the resonance peaks in Fig. 3 become narrower, while the detuning remains constant, making it harder for the comb teeth of this mode number to grow, thus requiring higher threshold power.

\section{\label{sec5}Conclusion}
In summary, we  develop a EPR entangled quantum frequency comb design platform based on a silicon nitride micro-ring resonator and employ the bipartite entanglement criterion to quantify the impact of the microresonator structure on the degree of entanglement. With our quantum frequency comb, we can provide at least 12 channels of quantum entangled states, making it an ideal solution for multi-channel quantum information networks. We  match different microcavity structures to various dispersions and conduct an in-depth analysis of the entanglement bandwidth and threshold power under both normal and anomalous dispersion. Additionally, we use one of the six  generated EPR entangled pairs to produce a frequency-dependent single-mode squeezed state, studying the relationship between the readout angle and the observation frequency to analyze the frequency-dependent squeezing. Our findings have potential applications in  quantum precision measurement.

\medskip
\textbf{} \par 

\medskip
\textbf{Acknowledgements} \par 
This work is supported by the Key-Area Research and Development Program of Guangdong Province (Grant No.2018B030325002), the National Natural Science Foundation of China (Grant No.62075129, 61975119), the Open Project Program of SJTU-Pinghu Institute of Intelligent Optoelectronics (Grant No.2022SPIOE204) and the Sichuan Provincial Key Laboratory of Microwave Photonics (Grant 2023-04).

\medskip

%
\bibliographystyle{MSP}
\bibliography{ref.bib}

\begin{thebibliography}{10}
\providecommand{\url}[1]{\texttt{#1}}
\providecommand{\urlprefix}{URL }

\bibitem{yu2020nature}
H.~Yu, et~al. (LIGO Scientific~Collaboration),
\newblock \emph{Nature} \textbf{2020}, \emph{583} 43.

\bibitem{acernese2020virgo}
F.~Acernese, et~al. (The Virgo~Collaboration),
\newblock \emph{Physical Review Letters} \textbf{2020}, \emph{125} 131101.

\bibitem{caves1981quantum}
C.~M. Caves,
\newblock \emph{Phys. Rev. D} \textbf{1981}, \emph{23} 1693.

\bibitem{walls1983squeezed}
D.~F. Walls,
\newblock \emph{Nature} \textbf{1983}, \emph{306} 141.

\bibitem{breitenbach1997measurement}
G.~Breitenbach, S.~Schiller, J.~Mlynek,
\newblock \emph{Nature} \textbf{1997}, \emph{387} 471.

\bibitem{schnabel2017squeezed}
R.~Schnabel,
\newblock \emph{Phys. Rep.} \textbf{2017}, \emph{684} 1.

\bibitem{ligo2011gravitational}
T.~L.~S. Collaboration,
\newblock \emph{Nat. Phys.} \textbf{2011}, \emph{7} 962.

\bibitem{ligo2013enhanced}
T.~L.~S. Collaboration,
\newblock \emph{Nature Photonics} \textbf{2013}, \emph{7} 613.

\bibitem{kimble2001conversion}
H.~J. Kimble, Y.~Levin, A.~B. Matsko, K.~S. Thorne, S.~P. Vyatchanin,
\newblock \emph{Physical Review D} \textbf{2001}, \emph{65} 022002.

\bibitem{chelkowski2005experimental}
S.~Chelkowski, \emph{et al.},
\newblock \emph{Physical Review A} \textbf{2005}, \emph{71} 013806.

\bibitem{khalili2010optimal}
F.~Y. Khalili,
\newblock \emph{Physical Review D} \textbf{2010}, \emph{81} 122002.

\bibitem{barsotti2018squeezed}
L.~Barsotti, J.~Harms, R.~Schnabel,
\newblock \emph{Reports on Progress in Physics} \textbf{2018}, \emph{82} 016905.

\bibitem{vyatchanin1995frequency}
S.~Vyatchanin, E.~Zubova,
\newblock \emph{Physics Letters A} \textbf{1995}, \emph{201} 269.

\bibitem{braginsky1980science}
V.~B. Braginsky, Y.~I. Vorontsov, K.~S. Thorne,
\newblock \emph{Science} \textbf{1980}, \emph{209} 547.

\bibitem{vasilakis2015natphys}
G.~Vasilakis, H.~Shen, K.~Jensen, M.~Balabas, D.~Salart, B.~Chen, E.~S. Polzik,
\newblock \emph{Nature Physics} \textbf{2015}, \emph{11} 389.

\bibitem{hertzberg2010natphys}
J.~B. Hertzberg, T.~Rocheleau, T.~Ndukum, M.~Savva, A.~A. Clerk, K.~C. Schwab,
\newblock \emph{Nature Physics} \textbf{2010}, \emph{6} 213.

\bibitem{suh2014science}
J.~Suh, A.~J. Weinstein, C.~U. Lei, E.~E. Wollman, S.~K. Steinke, P.~Meystre, A.~A. Clerk, K.~C. Schwab,
\newblock \emph{Science} \textbf{2014}, \emph{344} 1262.

\bibitem{shomroni2019natcomm}
I.~Shomroni, L.~Qiu, D.~Malz, A.~Nunnenkamp, T.~J. Kippenberg,
\newblock \emph{Nature Communications} \textbf{2019}, \emph{10} 2086.

\bibitem{chen2011genrelgrav}
Y.~Chen, S.~L. Danilishin, F.~Y. Khalili, H.~Müller-Ebhardt,
\newblock \emph{General Relativity and Gravitation} \textbf{2011}, \emph{43} 671.

\bibitem{ma2017proposal}
Y.~Ma, \emph{et al.},
\newblock \emph{Nature Physics} \textbf{2017}, \emph{13} 776.

\bibitem{yap2020generation}
M.~J. Yap, \emph{et al.},
\newblock \emph{Nature Photonics} \textbf{2020}, \emph{14} 102.

\bibitem{sudbeck2020demonstration}
J.~Südbeck, S.~Steinlechner, M.~Korobko, R.~Schnabel,
\newblock \emph{Nature Photonics} \textbf{2020}, \emph{14} 1.

\bibitem{vahlbruch2006coherent}
H.~Vahlbruch, \emph{et al.},
\newblock \emph{Physical Review Letters} \textbf{2006}, \emph{97} 011101.

\bibitem{chua2011backscatter}
S.~S.~Y. Chua, \emph{et al.},
\newblock \emph{Optics Letters} \textbf{2011}, \emph{36} 4680.

\bibitem{junker2022frequency}
J.~Junker, D.~Wilken, N.~Johny, D.~Steinmeyer, M.~Heurs,
\newblock \emph{Physical Review Letters} \textbf{2022}, \emph{129} 033602.

\bibitem{leonhardt1997measuring}
U.~Leonhardt,
\newblock \emph{Measuring the Quantum State of Light},
\newblock Cambridge University Press, Cambridge, England, \textbf{1997}.

\bibitem{hage2010towards}
B.~Hage, A.~Samblowski, R.~Schnabel,
\newblock \emph{Physical Review A} \textbf{2010}, \emph{81} 062301.

\bibitem{li2024platform}
N.~Li, B.~Ji, Y.~Shen, G.~He,
\newblock \emph{Physical Review Applied} \textbf{2024}, \emph{21} 054058.

\bibitem{dudley2006supercontinuum}
J.~M. Dudley, G.~Genty, S.~Coen,
\newblock \emph{Reviews of Modern Physics} \textbf{2006}, \emph{78} 1135.

\bibitem{shi2023entanglement}
H.~Shi, Z.~Chen, S.~E. Fraser, M.~Yu, Z.~Zhang, Q.~Zhuang,
\newblock \emph{npj Quantum Information} \textbf{2023}, \emph{9} 91.

\bibitem{godey2014stability}
C.~Godey, I.~V. Balakireva, A.~Coillet, Y.~K. Chembo,
\newblock \emph{Physical Review A} \textbf{2014}, \emph{89} 063814.

\bibitem{vernon2015spontaneous}
Z.~Vernon, J.~E. Sipe,
\newblock \emph{Physical Review A} \textbf{2015}, \emph{91} 053802.

\bibitem{vernon2015strongly}
Z.~Vernon, J.~E. Sipe,
\newblock \emph{Physical Review A} \textbf{2015}, \emph{92} 033840.

\bibitem{ji2024simultaneous}
B.~Ji, N.~Li, G.~He,
\newblock \emph{arXiv} \textbf{2024}, \emph{2406.16622}.

\bibitem{javid2023chipscale}
U.~A. Javid, R.~Lopez-Rios, J.~Ling, A.~Graf, J.~Staffa, Q.~Lin,
\newblock \emph{Nature Photonics} \textbf{2023}, \emph{17} 883.

\bibitem{zeng2024quantum}
H.~Zeng, Z.-Q. He, Y.-R. Fan, \emph{et al.},
\newblock \emph{Physical Review Letters} \textbf{2024}, \emph{132} 133603.

\bibitem{wang2004broadband}
Y.~Wang, W.~Zhang, Q.~Wang, X.~Feng, X.~Liu, J.~Peng,
\newblock \emph{Optics Letters} \textbf{2004}, \emph{29}, 8 889.

\bibitem{chembo2016quantum}
Y.~K. Chembo,
\newblock \emph{Physical Review A} \textbf{2016}, \emph{93} 033820.

\bibitem{whalen2016time}
S.~J. Whalen, H.~J. Carmichael,
\newblock \emph{Physical Review A} \textbf{2016}, \emph{93} 063820.

\bibitem{duan2000inseparability}
L.-M. Duan, G.~Giedke, J.~I. Cirac, P.~Zoller,
\newblock \emph{Physical Review Letters} \textbf{2000}, \emph{84} 2722.

\bibitem{wang2023ultrasensitive}
D.~Wang, Q.~Wang, Q.~Zhang, Y.~Li,
\newblock \emph{Journal of the Optical Society of America B} \textbf{2023}, \emph{40}, 3 March 2023.

\bibitem{gonzalez2017third}
C.~González-Arciniegas, N.~Treps, P.~Nussenzveig,
\newblock \emph{Optics Letters} \textbf{2017}, \emph{42} 4865.

\bibitem{ng2023quantum}
E.~Ng, \emph{et al.},
\newblock \emph{arXiv:2307.05464} \textbf{2023}.

\bibitem{vendromin2024quantum}
C.~Vendromin, \emph{et al.},
\newblock \emph{arXiv:2404.15563} \textbf{2024}.

\bibitem{chembo2010modal}
Y.~K. Chembo, N.~Yu,
\newblock \emph{Physical Review A} \textbf{2010}, \emph{82} 033801.

\bibitem{matsko2005optical}
A.~B. Matsko, A.~A. Savchenkov, D.~Strekalov, V.~S. Ilchenko, L.~Maleki,
\newblock \emph{Physical Review A} \textbf{2005}, \emph{71} 033804.

\end{thebibliography}


\begin{figure}[htb]
	\includegraphics[width=1\linewidth]{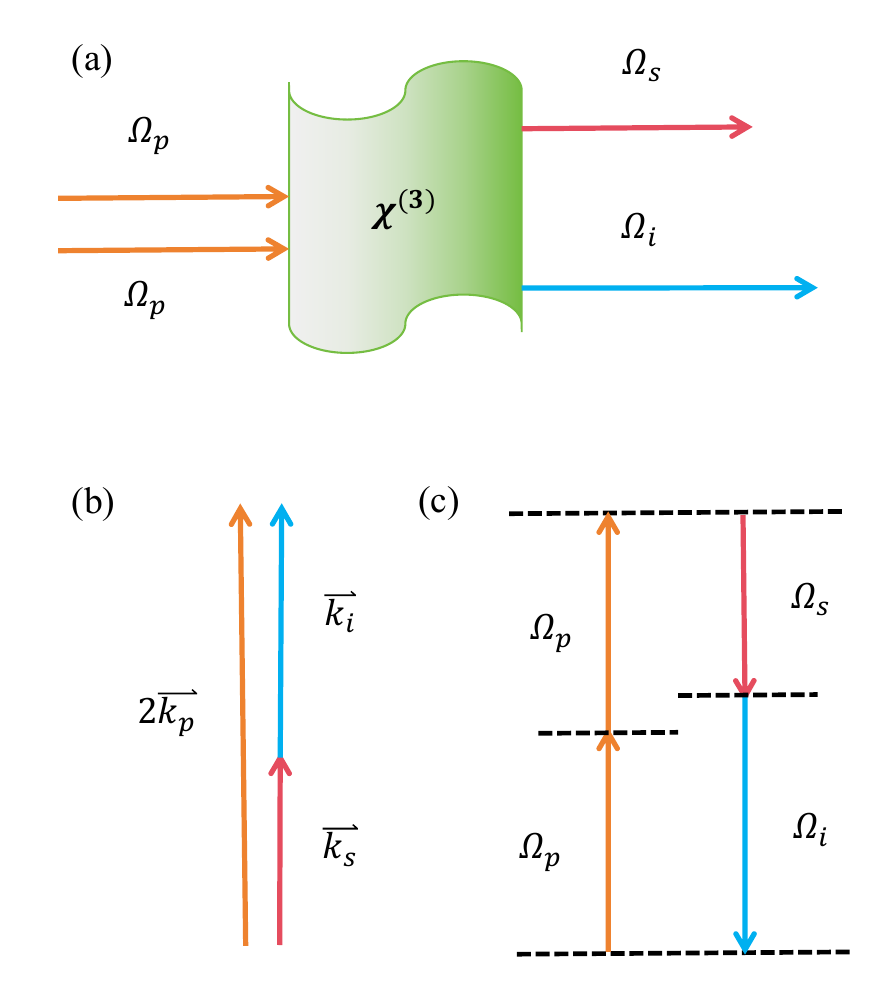}
	\caption{\label{f1} (a)A schematic depiction of the third-order nonlinear process in the generation of Kerr optical frequency combs.
 (b)The momentum conservation in four-wave mixing.
 (c)The energy conservation in four-wave mixing.}
\end{figure}
\begin{figure}[htb]
	\includegraphics[width=0.9\linewidth]{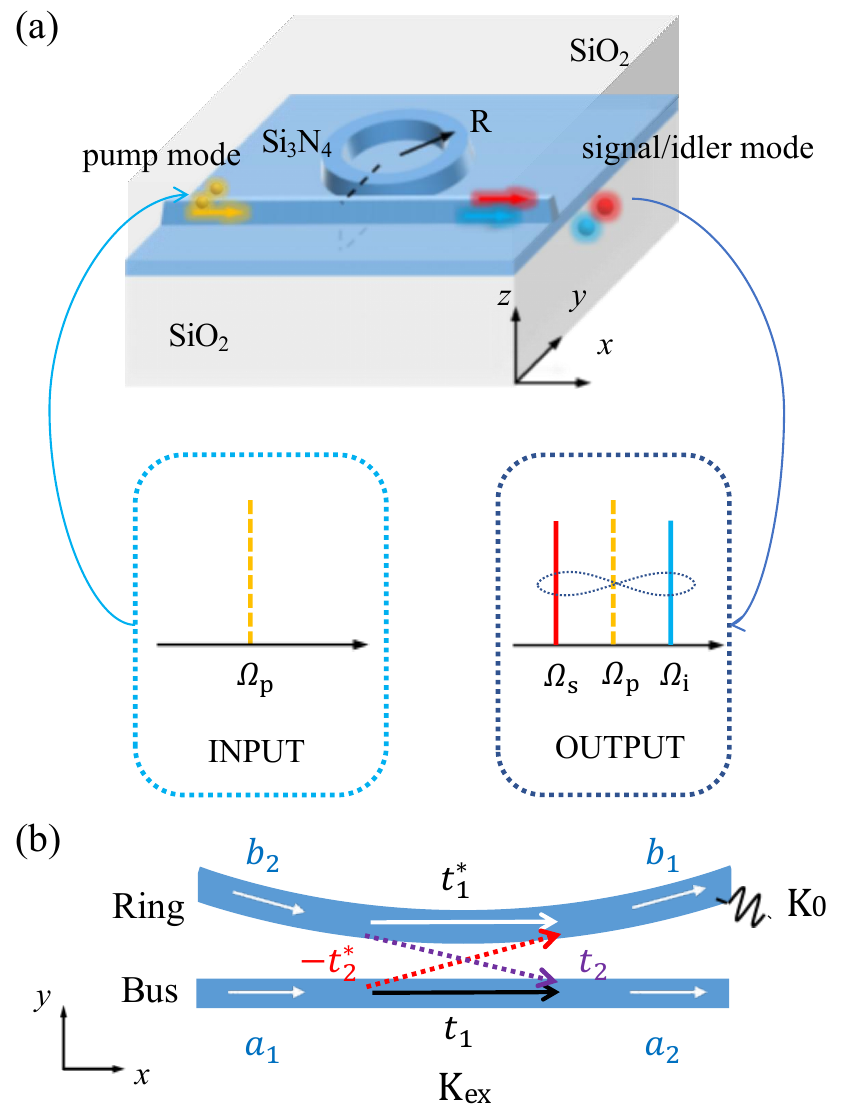}
	\caption{\label{f2} On-chip add-through microring resonator. (a) Three-dimensional scheme of the resonator. $R$ is the mean radius of the ring waveguide.  (b) Zoom-in view of the coupling region and its input-output schematic. $\kappa_0$, loss coupling rate; $\kappa_{ex}$, ring-bus coupling rate.The constants \( t_1 \) and \( t_2 \) satisfy $|t_1|^2+|t_2|^2=1$. Note that bus waveguides are not limited to be straight.}
\end{figure}
\begin{figure}[htb]
	\includegraphics[width=1\linewidth]{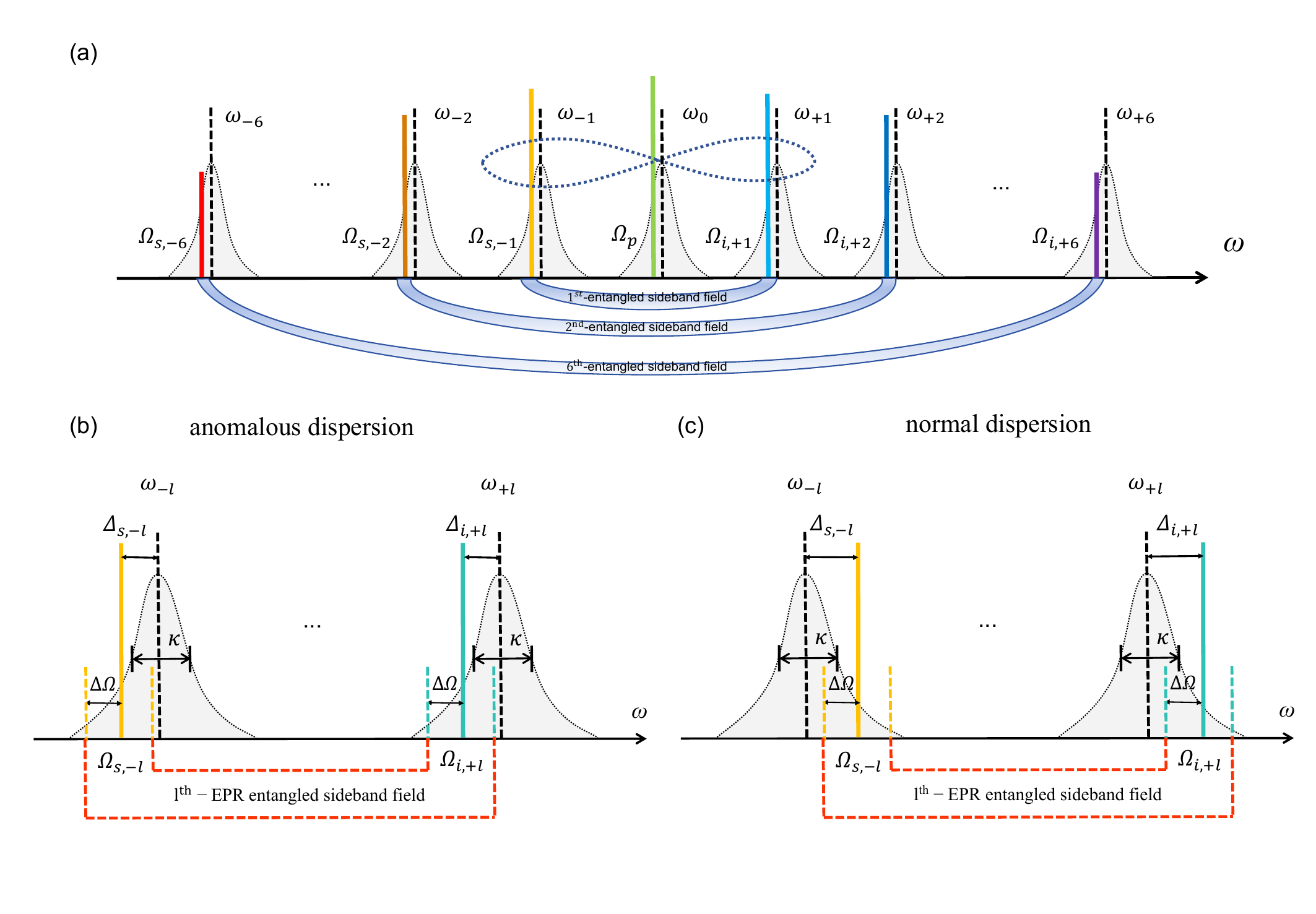}
	\caption{\label{f3} (a)Resonances under anomalous dispersion. Solid lines are perfectly equispaced, which represents the location of the output comb lines. Gray slashes represent the location of the resonances under anomalous dispersion.(b)An enlarged subgraph. In the case of mode 0, we know the relationship between the laser frequency $\Omega_0$ (i.e., $\Omega_p$), cold-resonance frequency $\omega_0$, loaded linewidth $\kappa$, and the normalized cold cavity pump detuning (red-detuned) $\sigma_c =\omega_0-\Omega_0$. As for mode $+l$, the normalized cold cavity detuning $\Delta_{i,+l}=\omega_{+l}-\Omega_{i,+l}$. Assume the considered modes share the same $\kappa$.(c)Enlarged subfigure for normal dispersion.}
\end{figure}
\begin{figure}[htb]
	\includegraphics[width=1.1\linewidth]{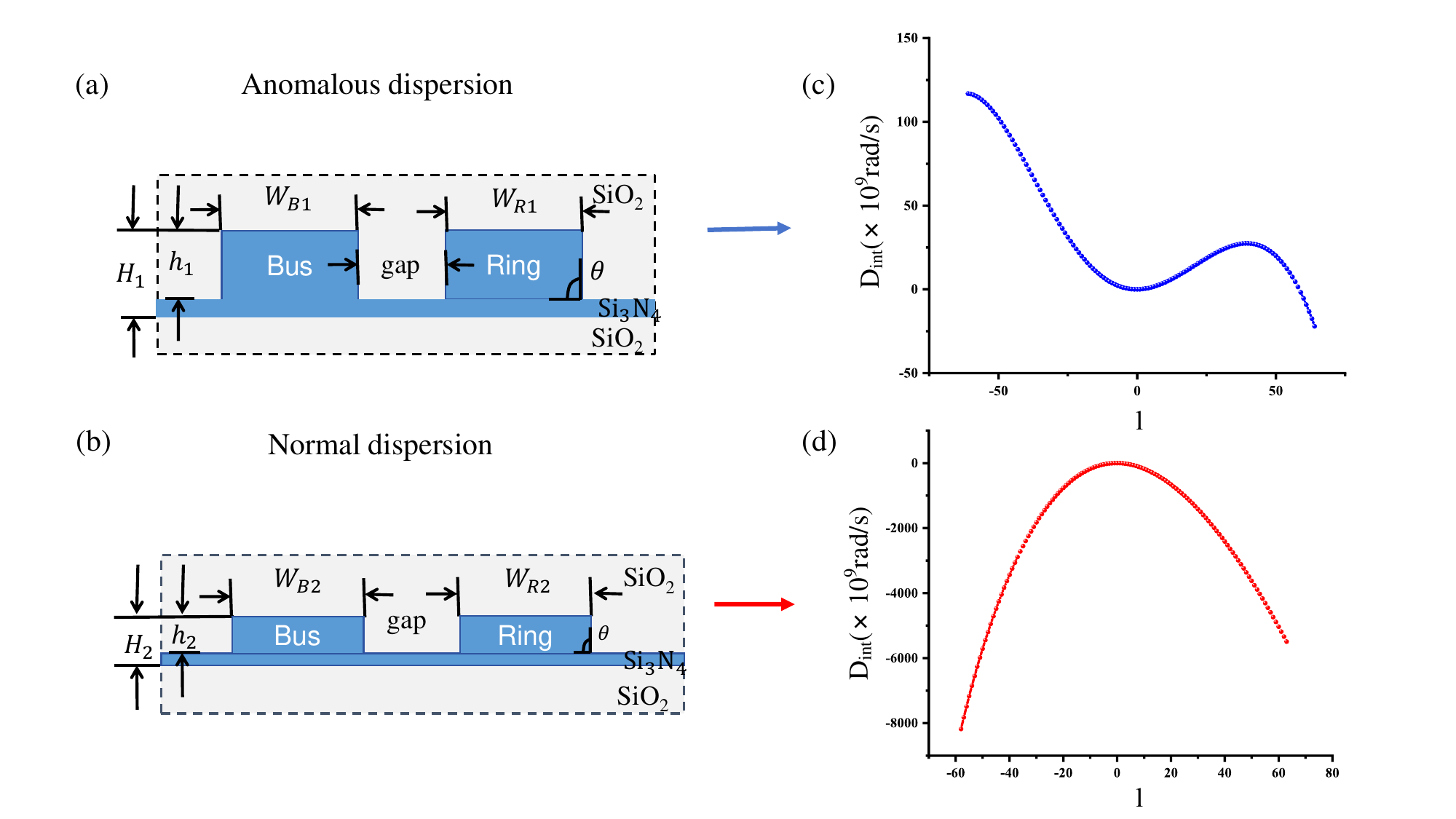}
	\caption{\label{f5} (a) Profile of rib waveguides under anomalous dispersion. (b) Profile of rib waveguides under normal dispersion. (c) Dispersion simulation curve under anomalous dispersion.Here, l represents the relative mode number. (d) Dispersion simulation curve under normal dispersion.}
\end{figure}
\begin{figure*}[htbp]
	\includegraphics[width=1\linewidth]{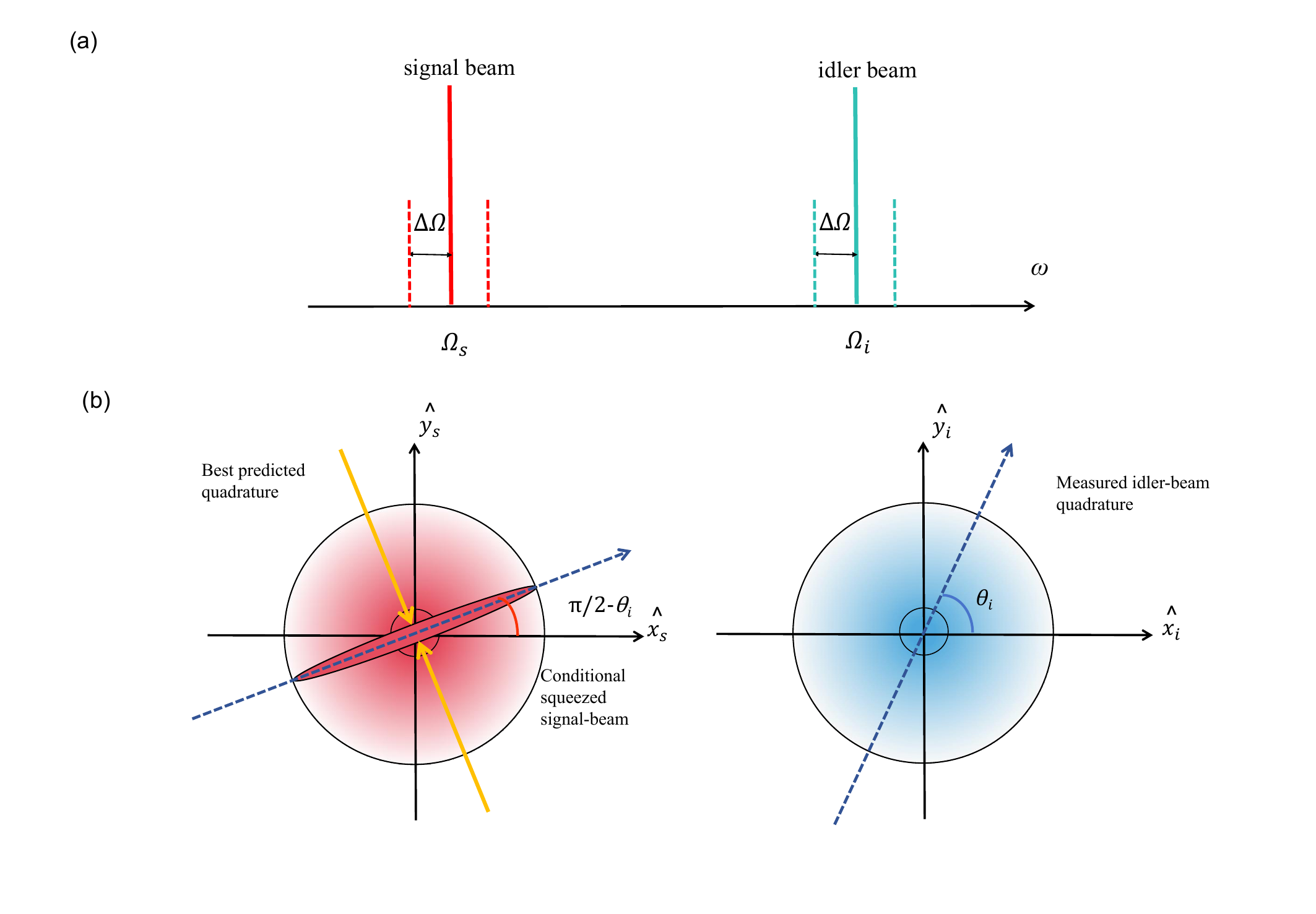}
	\caption{\label{below}  (a) The spectral decomposition of EPR-entangled beams, as referenced in Fig. 3. (b) Quantum statistics of the signal and idler beams. }
\end{figure*}
\begin{figure}[htbp]
	\includegraphics[width=0.8\linewidth]{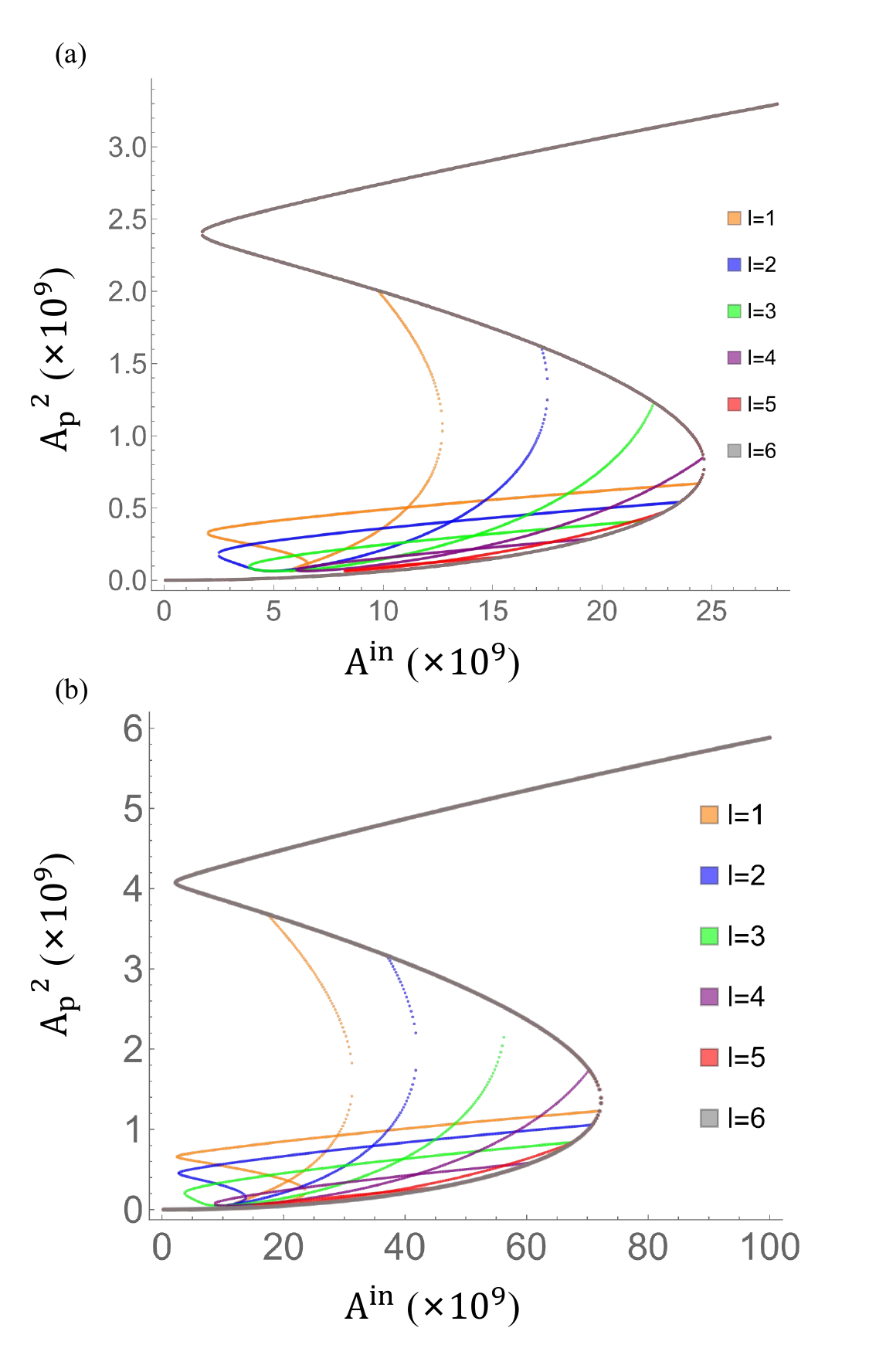}
	\caption{\label{below} The intracavity pump mode amplitude versus injected pump amplitude of our QFC. (a) Under anomalous dispersion($\sigma_c$ = 8 GHz). (b) Under normal dispersion ($\sigma_c$ = 18 GHz). }
\end{figure}

\begin{figure}[htbp]	\includegraphics[width=0.9\linewidth]{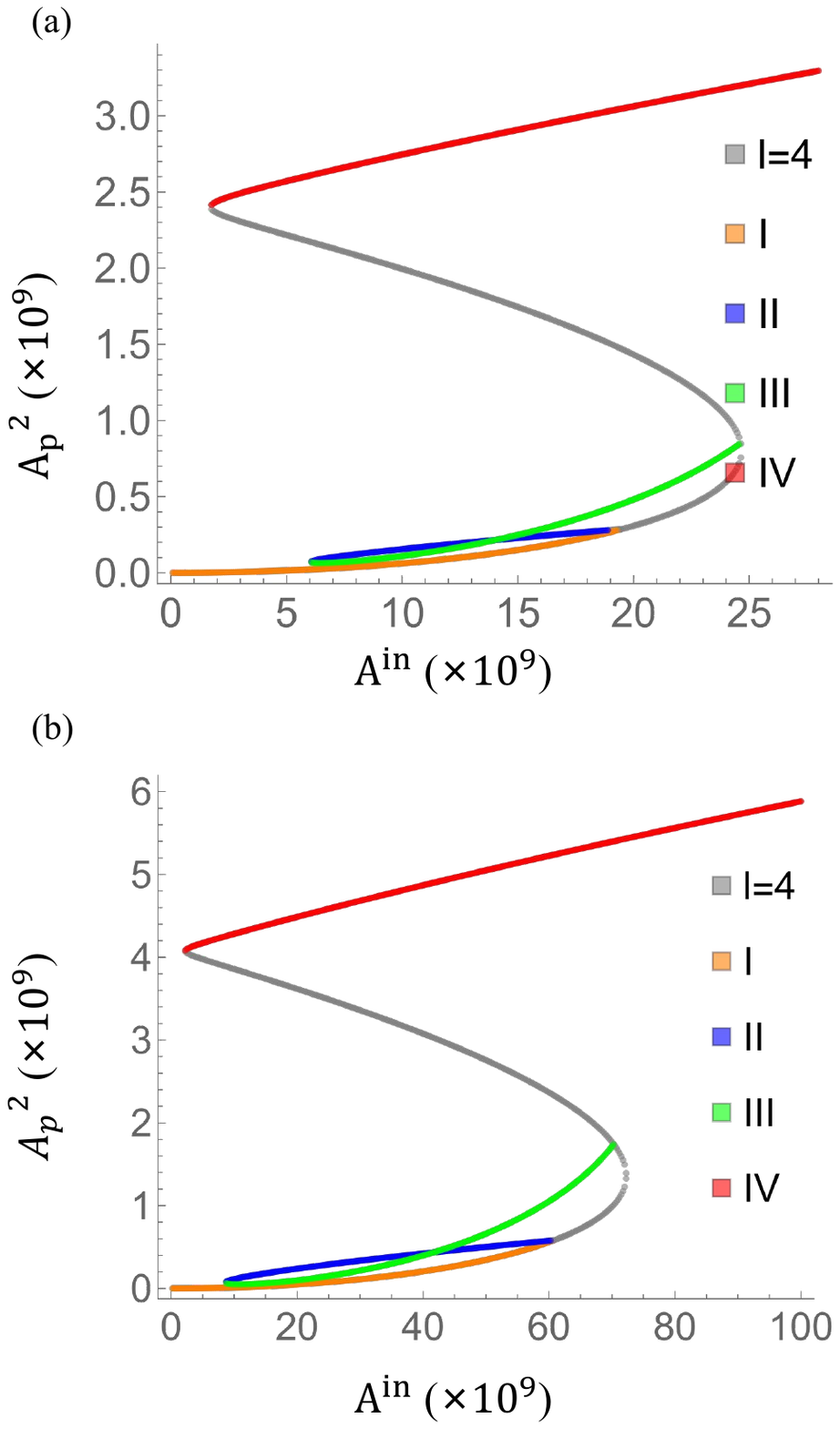}
	\caption{\label{below3} The intracavity pump mode amplitude versus injected pump amplitude of the fourth mode. (a) Under anomalous dispersion($\sigma_c$ = 8 GHz). (b) Under normal dispersion ($\sigma_c$ = 18 GHz).}
\end{figure}

\begin{figure*}[htbp]
	\includegraphics[width=1\linewidth]{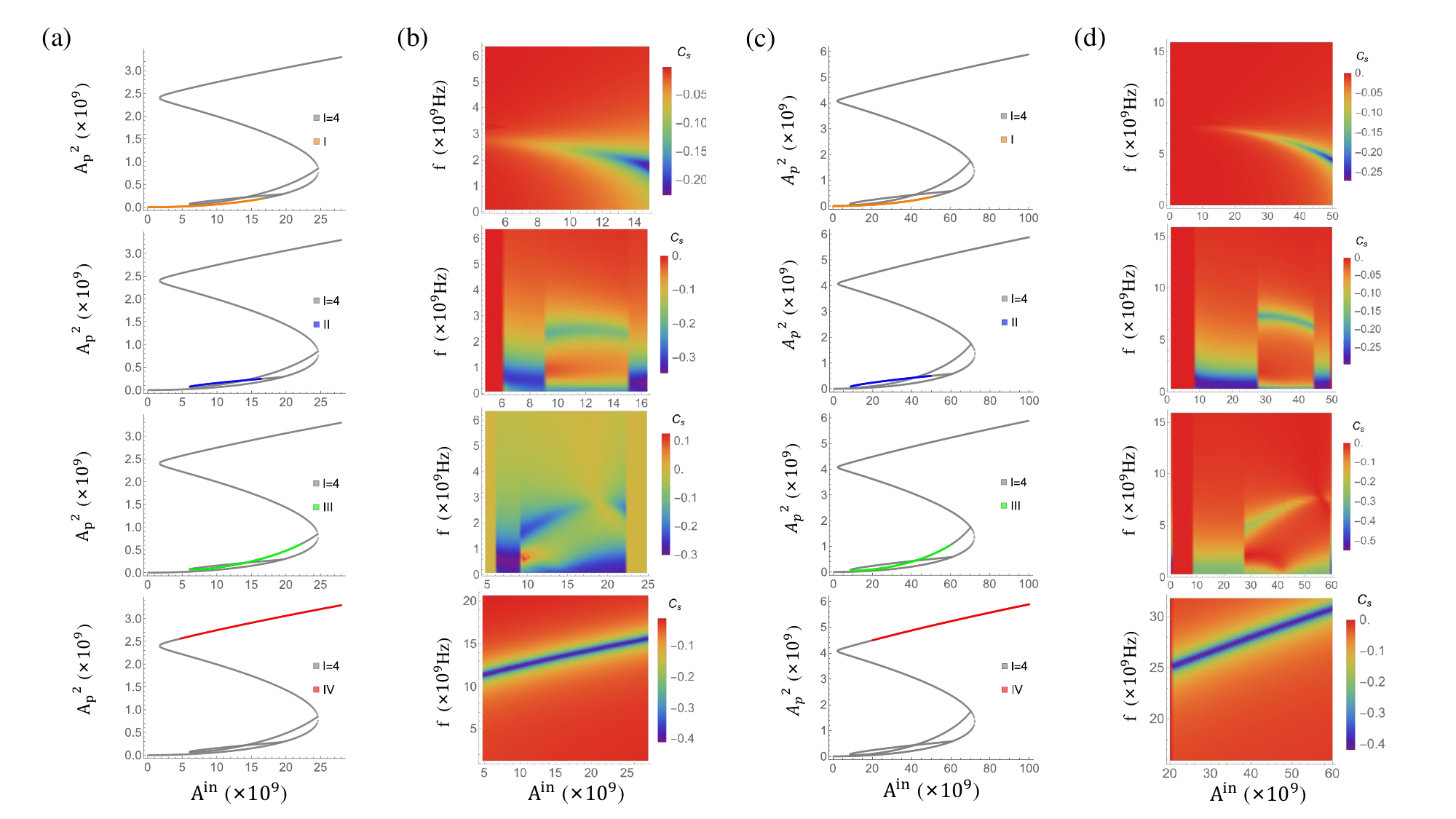}
	\caption{\label{below1} (a) The four stages of the fourth mode under anomalous dispersion. (b) The entanglement distribution in the regime of the injected pump amplitude $A^{in}$ and observation frequency f under anomalous dispersion($\sigma_c$ = 8GHz, $r$=1.222). (c) The four stages of the fourth mode under normal dispersion. (d) The entanglement distribution in the regime of the injected pump amplitude $A^{in}$ and observation frequency f under normal dispersion($\sigma_c$ = 18GHz, $r$=1.222). }
\end{figure*}
\begin{figure*}[htbp]
	\includegraphics[width=1\linewidth]{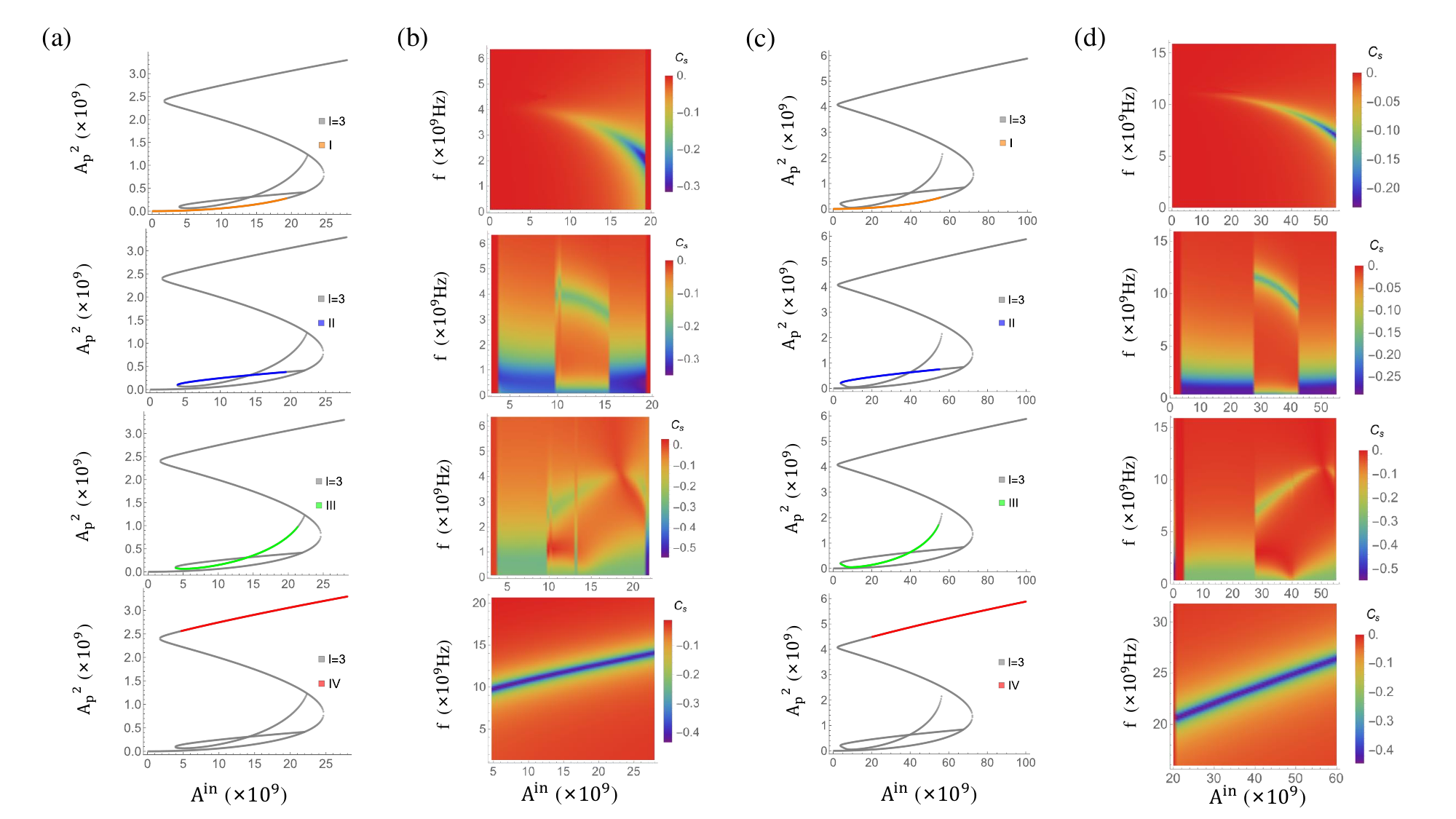}
	\caption{\label{below2} (a) The four stages of the third mode under anomalous dispersion. (b) The entanglement distribution in the regime of the injected pump amplitude $A^{in}$ and observation frequency f under anomalous dispersion($\sigma_c$ = 8GHz,$r$=1.222). (c) The four stages of the third mode under normal dispersion. (d) The entanglement distribution in the regime of the injected pump amplitude $A^{in}$ and observation frequency f under normal dispersion($\sigma_c$ = 18GHz, $r$ = 1.222). }
\end{figure*}
\begin{figure}[htbp]
	\includegraphics[width=0.8\linewidth]{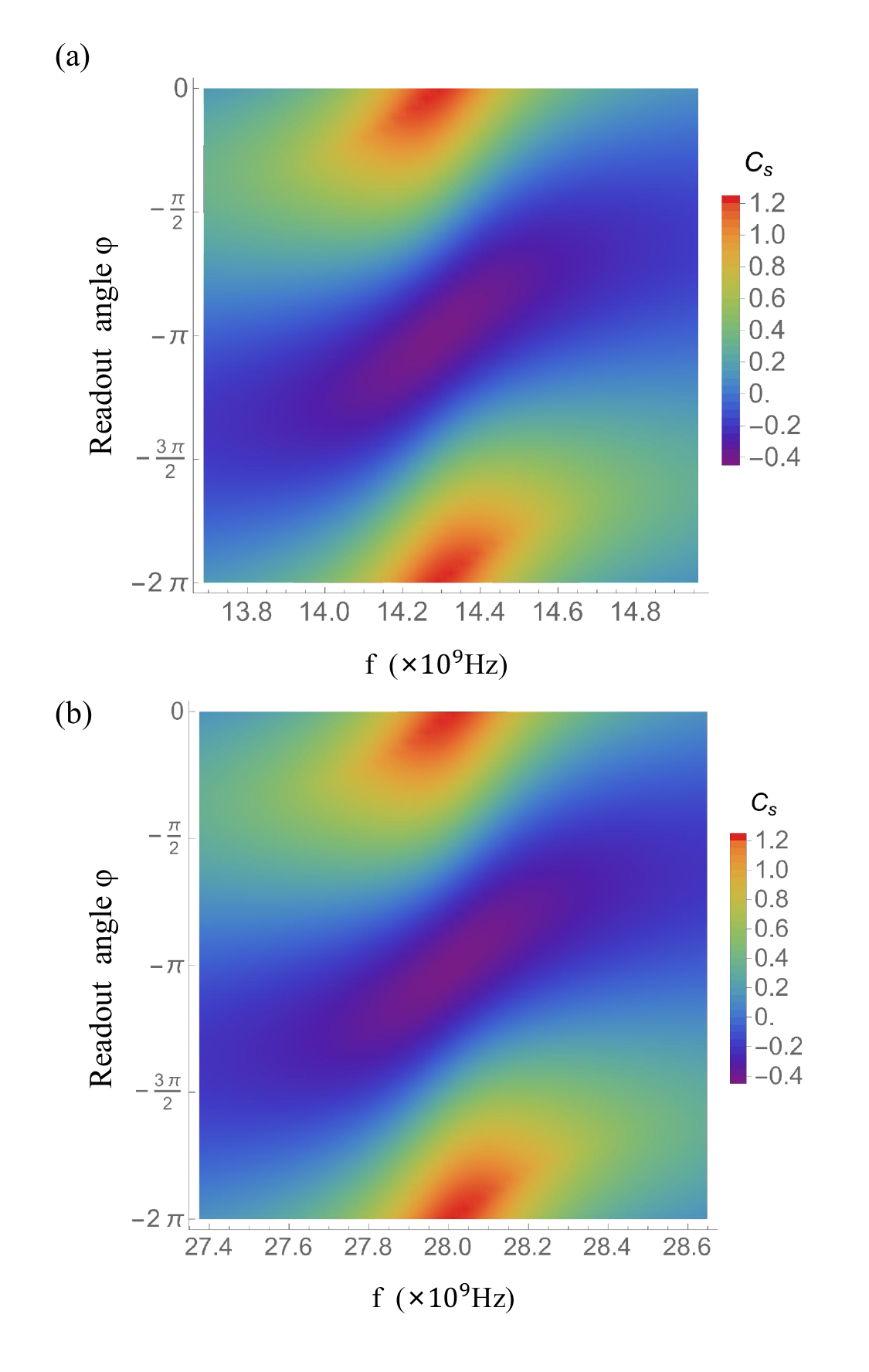}
	\caption{\label{below} The relationship diagram between the entanglement degree $C_s$, the observation frequency f (where f$=\frac{\Delta \Omega}{2\pi}$), and the readout angle, which shows frequency-dependent squeezing via Einstein–Podolsky–Rosen entanglement. (a) Under anomalous dispersion ($\sigma_c$ = 8 GHz, r=1.222,$A^{in}$ = $1\times10^{10}$ $\text{V/m}$). (b) Under normal dispersion ($\sigma_c$ = 18 GHz, r=1.222,$A^{in}$ = $4\times10^{10}$ $\text{V/m}$). }
\end{figure}
\begin{figure}[htbp]
	\includegraphics[width=0.8\linewidth]{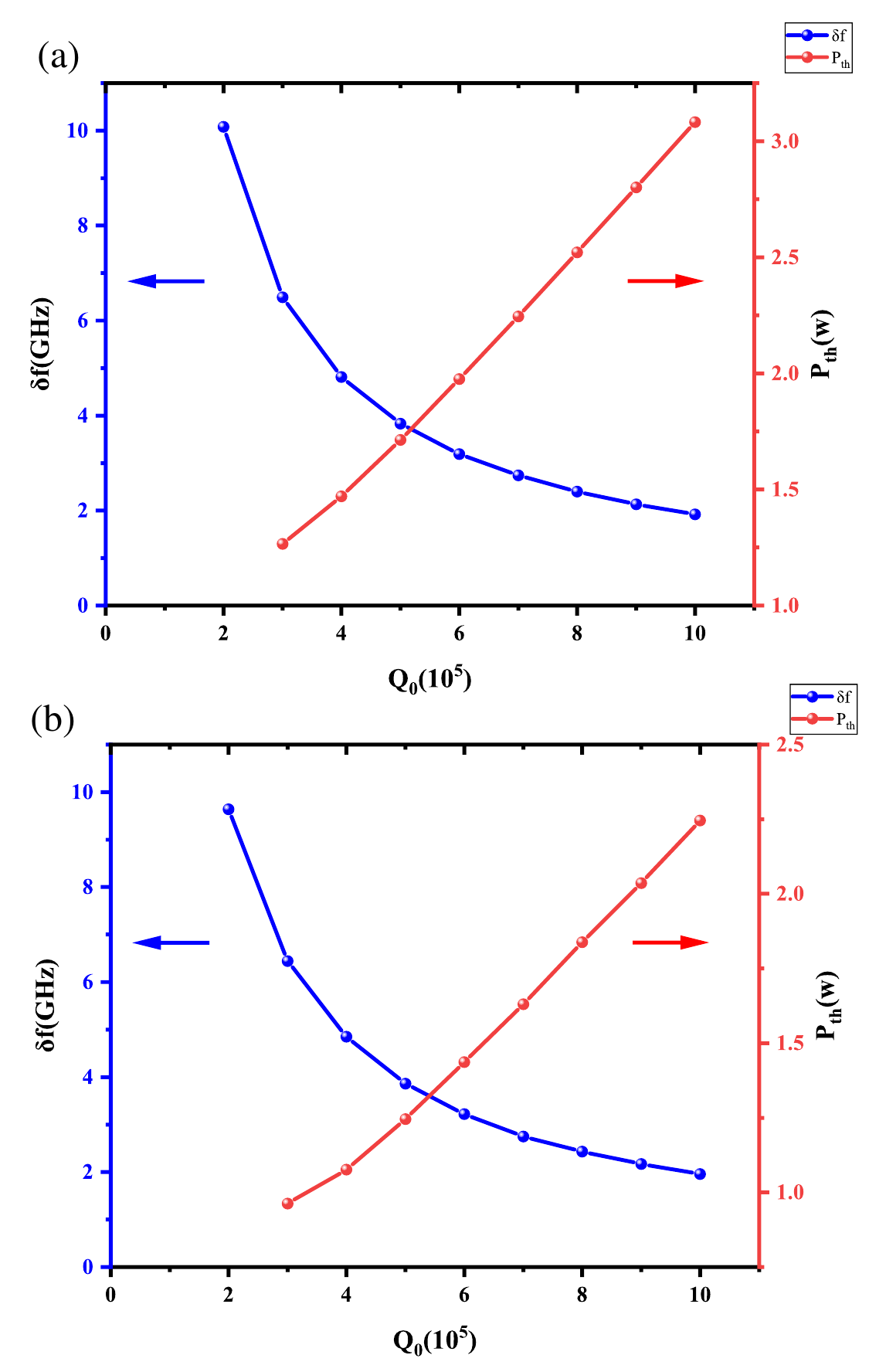}
	\caption{\label{below} The relationship curve between the intrinsic quality factor \(Q_0\), the entanglement bandwidth \(\delta f\), and the threshold power \(P_{\text{th}}\). (a) Under anomalous dispersion ($\sigma_c$ = 3 GHz, r=1.222). (b) Under normal dispersion ($\sigma_c$ = 3 GHz, r=1.222). }
\end{figure}


\begin{figure}
\textbf{Table of Contents}\\
\medskip
\includegraphics[width=55mm,height=50mm]{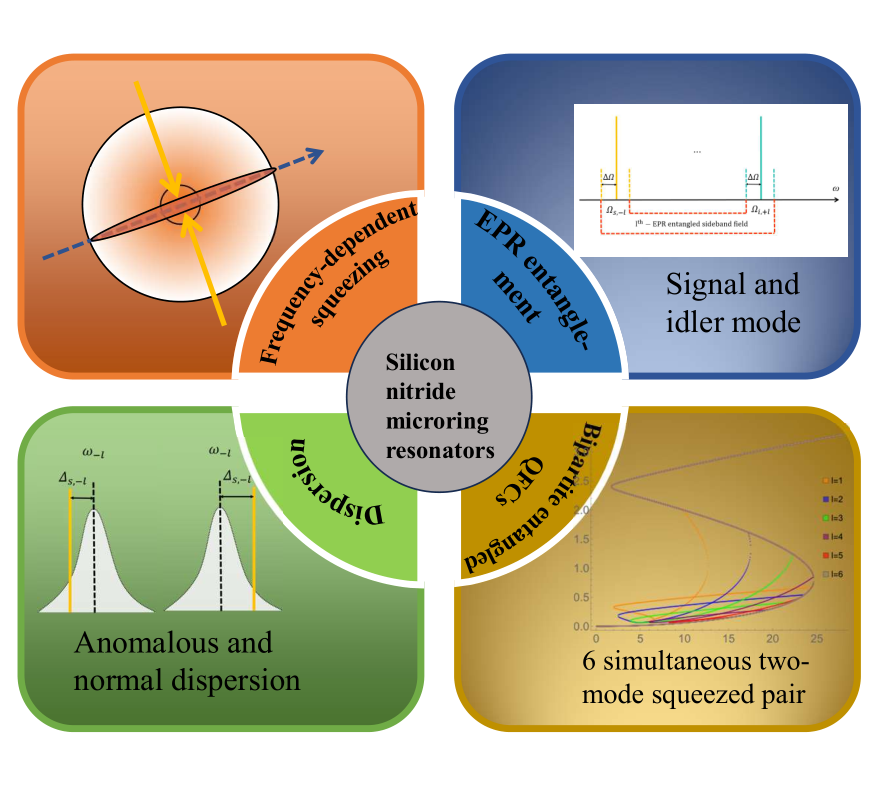} 
\medskip
\caption*{This study establishes  quantum optical frequency combs using a silicon nitride microring resonator, capable of supporting at least 6 simultaneous two-mode squeezed pairs. One of these EPR entangled pairs is used to generate a single-mode squeezed state, with an analysis of its frequency dependence. The entanglement characteristics under normal and anomalous dispersion conditions are also compared.}
\end{figure}

\end{document}